\documentclass[lettersize, journal]{IEEEtran}
\usepackage[backend=biber, url=false, eprint=false, doi=false, sorting=none]{biblatex}

\AtEveryBibitem{
\clearfield{urldate}
}
\AtEveryBibitem{\clearlist{language}}
\usepackage{graphicx}
\usepackage{hyperref}
\usepackage{xurl}
\usepackage{algorithm}
\usepackage{subcaption}
\usepackage{amsmath}
\usepackage{makecell}
\allowdisplaybreaks
\usepackage{amsfonts}
\usepackage{pifont}
\newcommand{\cmark}{\ding{51}}
\newcommand{\xmark}{\ding{55}}
\usepackage{xcolor}
\usepackage{multirow}
\usepackage{multicol}
\usepackage{amssymb}
\usepackage{amsthm}
\usepackage{algpseudocode}
\usepackage{accents}
\newtheorem{theorem}{Theorem}[section]
\newtheorem{example}{Example}
\newtheorem{proposition}{Proposition}[section]
\newtheorem{definition}{Definition}[section]
\newtheorem{assumption}{Assumption}[section]

\theoremstyle{definition}
\newtheorem{case}{Case}

\DeclareMathOperator*{\argmin}{arg\,min}
\usepackage{pdfpages}
\usepackage{tabularx}
\usepackage{array}
\newcolumntype{Y}{>{\centering\arraybackslash}X}
\usepackage{comment}
\usepackage{booktabs}
\usepackage{siunitx}
\definecolorset{RGB}{}{}%
{sycamoregreen,0,148,68}
\definecolorset{RGB}{}{}{bluee, 10,10,160}

\ifCLASSINFOpdf
\else
\fi
\begin{document}
%
\title{Uncertainty and Autarky: Cooperative Game Theory\\ for Stable Local Energy Market Partitioning}
%
%
%

\author{Saurabh Vaishampayan,\ Maryam Kamgarpour ~\IEEEmembership{Senior Member,~IEEE}
\thanks{We gratefully acknowledge funding under ``UrbanTwin'' with the financial support of the ETH-Domain Joint Initiative program in the Strategic Area Energy, Climate and Sustainable Environment. }
\thanks{Saurabh Vaishampayan and Maryam Kamgarpour are with the SYCAMORE Lab EPFL, 1015 Lausanne, Switzerland (email: saurabh.vaishampayan@epfl.ch, maryam.kamgarpour@epfl.ch).}}
\maketitle

\begin{abstract}

Local energy markets empower prosumers in distribution grids to form coalitions for collective self-consumption. An open question is to analyze the scale and composition of local energy market coalitions formed by strategic prosumers in distribution grids. This analysis must account for grid constraints, stochasticity of load and generation, as well as the interaction between possibly multiple local energy markets in the distribution grid. In this work, we present a cooperative game theoretic framework to study distribution grid partitioning into local energy markets under uncertain prosumption, grid constraints, and coalitional externalities. We formulate the optimal stable partitioning problem to balance the interests of the grid operator with that of strategic prosumers. Under deterministic load and generation, we show that the largest market coalition is the optimal stable partition. Under high levels of grid congestion, we show that individual self-consumption corresponds to the optimal stable partition. For the general case of stochastic prosumption and moderate grid congestion levels, we provide an algorithm to evaluate the optimal stable partition. We validate our algorithm and theory using numerical experiments on benchmark and real world distribution grids. Our results help in understanding the impact of prosumption uncertainty and grid constraints on coalition formation.

\end{abstract}
\begin{IEEEkeywords}
Uncertainty, Network constraints, Cooperative game theory, Local energy markets, Distribution grid partitioning
\end{IEEEkeywords}
\IEEEpeerreviewmaketitle

\section{Introduction}

%
%
%
%
\subsection{Motivation and related work}

\IEEEPARstart{R}{e}cent legislations \cite{le_conseil_federal_suisse_feuille_nodate,parlamento_italiano_act_2020} empower \emph{prosumers} in distribution grids to form Local Energy Markets (LEMs) for collective self-consumption. However, the scale and composition of the LEMs that prosumers would form remains unclear. In a large distribution grid, an LEM could range anywhere from a single prosumer self-consuming individually, to collective self-consumption by prosumers in a neighborhood, to the extreme case of the entire distribution grid as a single LEM. In the general case, a distribution grid may have multiple such LEM \emph{coalitions}, so an open question is which grid \emph{partitions} emerge when prosumers are empowered to form LEM \emph{coalitions}.

There are two important considerations that govern the composition of LEMs in grid partitions: the tradeoffs related to the scale of collective self-consumption, and the perspectives of the stakeholders in the grid regarding these tradeoffs. 

The first consideration is the tradeoffs related to the scale of collective self-consumption. Forming smaller LEMs for self-consumption (analogous to the concept of \emph{autarky}) may result in reduced power flows and grid usage, which can minimize the risk of line overloading and voltage violations in constrained grids under uncertain load and generation \cite{dudjak_impact_2021, dimovski_impact_2023}. On the other hand, such localized self-consumption may come at the expense of economic efficiency by compromising on the economic gains from energy trading, which is achieved in larger LEMs \cite{schmitt_how_2022}. Thus, the scale and composition of LEMs in grid partitioning is governed by the tradeoffs between risks associated with network constraint violations under uncertain load and generation, and economically efficient dispatch.

The second consideration in grid partitioning is accounting for the perspectives of the different stakeholders in the grid, who may evaluate the above mentioned tradeoffs differently. There are two types of stakeholders: \textbf{1. The Distribution System Operator (DSO)}, who manages the distribution grid and cares about aggregate costs for the grid; and \textbf{2. Prosumers}, who actually have the power to form LEMs, and may only care about local costs. The objectives of the DSO and the prosumers need not always align, and the grid partitions favoured by the two may be different\footnote{As we show later in Section \ref{sec:partitioningoutcomes} (Examples \ref{example:1}, \ref{example:2})}. Thus, an important step for optimal partitioning is the choice of the right solution concept accounting for these two perspectives. We first review the relevant literature for each of these perspectives.



\textbf{1. Grid partitioning from the DSO's perspective}. Here the optimal partitioning problem is formulated to minimize the aggregate costs for the whole grid. A number of papers in the literature have tackled this question. The authors \cite{barani_optimal_2019} first formulated the distribution grid partitioning problem in the context of microgrids. They jointly analyzed installation of distributed energy resources with grid partitioning into microgrids, accounting for voltage and thermal limits, and balancing constraints. The works \cite{osama_planning_2020,biswas_chance-constrained_2021} optimize partitioning accounting for the islanding capability of the partitioned microgrids. In the work \cite{wang_distributed_2025} optimal grid partitioning is studied jointly alongside distributed optimization for energy dispatch.

An important shortcoming of works in this strand of literature is that they assume that a centralized entity/DSO can enforce the optimal grid partition. Current local energy market legislations in various countries \cite{le_conseil_federal_suisse_feuille_nodate,parlamento_italiano_act_2020} give prosumers the power to form local energy markets. The DSO may only set energy tariffs for the LEMs in the distribution grid to recover its operating costs. For a given partition and energy tariffs, if a group of prosumers find that they can obtain lower costs by forming alternative LEM coalitions, they may deviate. This may result in a different partition being formed. This brings us to the perspective of prosumers in analyzing LEM partitioning.

To incorporate \textbf{2. prosumers' perspective} in LEM formation, game theory is typically used in the literature to model decision making by prosumers who may have differing and interdependent objectives. Non-cooperative game theory models settings when players take decisions unilaterally, and has previously been used in the context of local energy dispatch \cite{chen_peer--peer_2020,paudel_peer--peer_2019}. Cooperative game theory models settings when players coordinate to form coalitions and distribute the collective costs. In the context of grid partitioning, the main aspect of prosumers' decision making is the formation of LEM coalitions, thus cooperative game theory is a more suitable tool to study this problem.

Works using cooperative game theory to model LEM formation study \emph{coalition stability}, that is, whether a group of prosumers can sustainably form a coalition against the threat of alternative coalition formation by prosumers \cite{shapley_game_1973}. Here one calculates the total cost faced by prosumers for every possible LEM coalition. Then, allocation of this cost to members of the LEM coalition is analyzed. If cost allocations exist that prevent any subset of prosumers from seeking alternative coalitions, such an LEM is called a \emph{stable coalition}.

In the context of the energy dispatch problem, coalition stability was investigated by the authors \cite{han_incentivizing_2019, valencia_zuluaga_stable_2025, huang_uncertainty-aware_2024}, in the absence of power flow constraints. For distribution grids with AC power flow constraints, dynamic operating envelopes have been used to study fair allocations in unbalanced networks \cite{azim_dynamic_2024}, and hosting capacity determination \cite{azim_network_2025}. Coalition stability has also been studied in the context of installation and investment in new distributed energy resources \cite{fleischhacker_stabilizing_2022, fioriti_fair_2025}.

The previous literature on cooperative game theory for stable LEM coalition/partition formation formation suffers from two main drawbacks. Firstly, works either neglect realistic power flow models and constraints (\cite{han_incentivizing_2019,valencia_zuluaga_stable_2025, huang_uncertainty-aware_2024, fleischhacker_stabilizing_2022, fioriti_fair_2025, saad_coalitional_2011}) or prosumption uncertainty (\cite{valencia_zuluaga_stable_2025,azim_dynamic_2024, fleischhacker_stabilizing_2022, fioriti_fair_2025, saad_coalitional_2011}). The question on how both grid constraints and uncertainty affect LEM size and composition remains unclear.

Secondly, previous literature ignores the presence of \emph{coalitional externalities} in LEM formation. For any LEM in the distribution grids, coalition costs may not only depend on the composition of prosumers internal to it, but also how LEMs in the grid external to the given coalition are organized. This arises because power flow constraints result in the coupling of prosumers' energy dispatch, as well as the voltages and power flows across the distribution grid\footnote{see Section \ref{sec:coalitioncostssection} and Fig. \ref{fig:costexternalities} for a stylized example, and Appendix \ref{app:simulationslausanne} for a numerical example in a real grid}. This effect of cost coupling between LEM coalitions is known as \emph{coalitional externalities} \cite{thrall_nperson_1963, koczy_partition_2018}. Thus, a key assumption used in the previously mentioned works, that the coalition costs only depend on internal members and are independent of external prosumers, can no longer be justified.

Under coalitional externalities, to calculate its deviation costs, a deviating coalition must also anticipate the configuration of coalitions in the remainder grid following its deviation. In general, the grid can have multiple coalitions interacting with each other via externalities. Thus, the notion of coalition stability is extended to \emph{partition stability}. Here, each partition may consists of multiple coalitions, and in the case of stable partitions, no group of prosumers has the incentive to deviate from its existing coalition in the partition.

Research on partition stability under coalitional externalities in electric grids is limited. In the context of energy communities, the stable partitioning problem was used by the authors \cite{abada_unintended_2020} to study energy community formation in grids with shared infrastructure costs. In the work \cite{csercsik_efficiency_2017}, stable partitioning of transmission grids into balancing groups was studied using partition function form games \cite{thrall_nperson_1963}. However, both works neglect the presence of uncertain prosumption and the risk of network constraint violations.
\begin{table}[h!]
    \centering
    \begin{tabular}{|c|c|c|c|c|c|c|c|}
        \hline
        \makecell{Factors\\in LEM\\partitioning} & \makecell{\cite{barani_optimal_2019}\\ \cite{osama_planning_2020}\\\cite{biswas_chance-constrained_2021}\\ \cite{wang_distributed_2025}}& \makecell{\cite{valencia_zuluaga_stable_2025}\\\cite{saad_coalitional_2011}}& \makecell{\cite{han_incentivizing_2019}\\ \cite{huang_uncertainty-aware_2024}}& \makecell{\cite{azim_dynamic_2024}\\ \cite{azim_network_2025}}&\cite{abada_unintended_2020}&\cite{csercsik_efficiency_2017}&\makecell{\textbf{Our}\\ \textbf{work}}\\
        \hline
        \makecell{DSO's\\ perspective} & \cmark& \xmark& \xmark& \makecell{Partially\\(envelope\\setting)}& \cmark& \xmark&\cmark\\
        \hline
        \makecell{Prosumers'\\ perspective} & \xmark& \cmark& \cmark& \cmark& \cmark& \cmark&\cmark\\
        \hline
        \makecell{Prosumption\\uncertainty} & \cmark&\xmark & \cmark& \xmark& \xmark& \xmark&\cmark\\
        \hline
        \makecell{Grid\\constraints} &\cmark &\xmark &\xmark &\cmark & \xmark& \cmark&\cmark\\
        \hline
        \makecell{Coalitional\\externalities} & n/a& \xmark& \xmark& \xmark& \cmark& \cmark&\cmark\\
        \hline
    \end{tabular}
    \caption{Summary of literature on grid partitioning}
    \label{tab:litreview}
\end{table}

Table \ref{tab:litreview} summarizes various solutions outlined in literature.



\subsection{Contributions}
Our main contribution is the development of a cooperative game theoretic framework and solution concept called the optimal stable partition to balance the DSO's and prosumers' objectives; under grid constraints, prosumption uncertainty, and coalitional externalities. We provide analytical solutions for the optimal stable partition for the case of perfect forecasts (Theorem \ref{thm:stabilitystrictselfconsum}), and for the case congested grids (Proposition \ref{prop:stabilitycongested}). An algorithm (Algorithm \ref{alg:alg1}) is provided to calculate the optimal stable partition in the general case, which is validated on benchmark and real world grids.

\subsection{Organization}
The rest of the paper is organized as follows. Section \ref{sec:preliminaries} provides the preliminaries related to graph partitioning and power flow. In Section \ref{sec:probform}, we formulate a two-stage model to analyze total operating costs in a grid partition under uncertain prosumption. The model for coalition costs, that is the cost faced by individual LEM coalitions in a grid partition, is presented in Section \ref{sec:coalitioncostssection}. Based on these cost models, the partitioning preferences of the DSO and the prosumers are formulated in Section \ref{sec:partitioningoutcomes}. Following this, the optimal stable partitioning problem is formulated in Section \ref{sec:optimalstablepartition} to balance these preferences. Here, we provide an analytical solution for the case of perfect forecasts of prosumption (Theorem \ref{thm:stabilitystrictselfconsum}), and under fully congested grids (Proposition \ref{prop:stabilitycongested}). An algorithm (Algorithm \ref{alg:alg1}) is provided to calculate the solution under imperfect forecasts and moderate grid congestion. Section \ref{sec:numerics} validates the theory via simulations in benchmark and real world distribution grids, and conclusions given in Section \ref{sec:concl}. 


\section{Preliminaries}
\label{sec:preliminaries}
\subsection{Distribution grid partitions}
We consider a radial distribution grid graph $\mathcal{G} = (\mathcal{N}_{\mathcal{G}},\mathcal{E}_{\mathcal{G}})$ with nodes (buses) $\mathcal{N}_{\mathcal{G}}$ and edges (lines) $\mathcal{E}_{\mathcal{G}}$. Denote the Point-of-Common-Coupling of the distribution grid as $PCC(\mathcal{G})$. Denote the set of leaf nodes as $\mathcal{M}_{\mathcal{G}}\subset \mathcal{N}_{\mathcal{G}}$, with each prosumer assumed to be located at a unique leaf node. 

Let $\mathcal{S}(\mathcal{G})$ denote the set of connected subgraphs of $\mathcal{G}$. In this work, we only consider LEMs that are connected subgraphs of $\mathcal{G}$, hence, any LEM $F$ is an element of the set $\mathcal{S}(\mathcal{G})$.

For any LEM coalition $F$, denote its associated subgraph as $(\mathcal{N}_F, \mathcal{E}_F)$, where $\mathcal{N}_F, \mathcal{E}_F$ denote the nodes and internal edges. This LEM coalition hence contains the prosumer set given by $\mathcal{M}_F = \mathcal{N}_{F}\cap \mathcal{M}_{\mathcal{G}}$. The boundary nodes for LEM $F$, representing the nodes that have lines connecting to the rest of the distribution grid, are denoted by $\mathcal{N}_F^b$, and the corresponding boundary edges by $\mathcal{E}_F^{b}$. The internal nodes $\mathcal{N}_{F}\setminus\mathcal{N}_F^b$ are denoted as $\mathcal{N}_F^{int}$.

\begin{figure}[h!]
    \centering
    \includegraphics[width=0.49\linewidth]{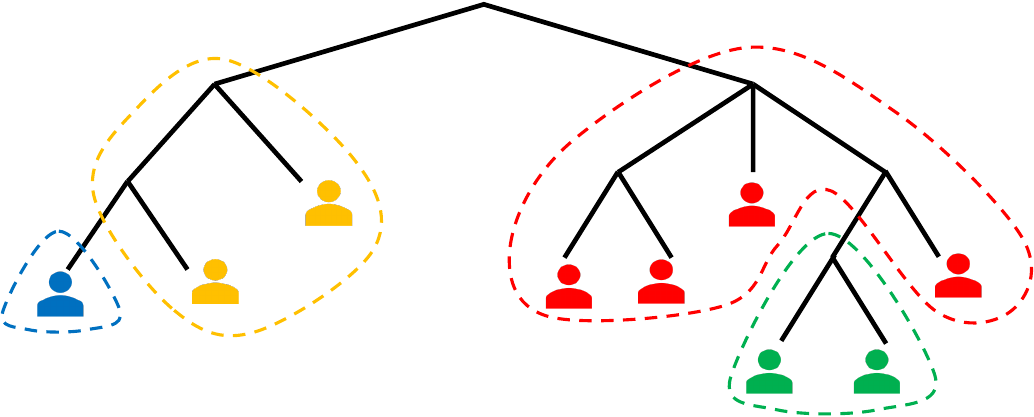}
    \caption{An example of a partition of a radial distribution grid}
    \label{fig:examplepartition}
\end{figure}

A distribution grid partition is then defined below as
\begin{definition}
    (Grid partition) Let $P=\{\{F_l\}_{l=1}^{L}\}$ denote a collection of connected subgraphs of the distribution grid $\mathcal{G}$. $P$ is a distribution grid partition if and only if
    \begin{enumerate}
        \item $\mathcal{E}_{F_i}\cap \mathcal{E}_{F_j} = \emptyset\ \forall i,j\in \{1,\cdots L\}, i\neq j$
        \item $\cup_{i=1}^{L} \mathcal{M}_{F_i} = \mathcal{M}_{\mathcal{G}}$
    \end{enumerate}
    \label{def:partition}
    with the set of all possible partitions of the distribution grid graph $\mathcal{G}$ denoted as $\mathcal{P}(\mathcal{G})$.
\end{definition}
Fig \ref{fig:examplepartition} illustrates a possible grid partition under Def \ref{def:partition}. 

\subsection{Power flow model}
Consider a time horizon $\mathcal{T} = \{1,2,\cdots T\}$. Let each prosumer node $m\in\mathcal{M}_{\mathcal{G}}$ have a source of stochastic inelastic prosumption and a source of controllable flexibility. The former, for time $t\in\mathcal{T}$, is denoted as $U_m^t$, with $U_m^t\in\mathbb{C}$ consisting of active and reactive power injections. The flexibility dispatch for the latter is denoted as $u_m^t\in \mathbb{C}$, with the net power injection given as the sum $U_m^t+u_m^t$. 

For brevity, denote the concatenation of any scalar $\zeta_n^t$ over the horizon $\mathcal{T}$ as $\boldsymbol{\zeta}_n$, and the concatenation over $n\in\mathcal{N}, t\in\mathcal{T}$ as $\boldsymbol{\zeta}$. Let $\boldsymbol{S}_{i\rightarrow j}\in\mathbb{C}^{T}$ denote the complex power flows in the edge $i\rightarrow j$. Let $\boldsymbol{v}_n\in\mathbb{C}^T$ denote the vector of (squared) nodal voltages at node $n$, with $\boldsymbol{\delta v}_n = \boldsymbol{v}_n-v_{ref}$ denoting the (squared) voltage deviations with respect to a reference $v_{ref}$. 

In our work, we use the Linear DistFlow equations \cite{baran_optimal_1989} to model AC power flow, which are adapted as
\begin{subequations}
    \begin{align}
         &\boldsymbol{\delta v}_{A_n} = \boldsymbol{\delta v}_n-2\mathrm{Re}(z_n^*\boldsymbol{S}_{n\rightarrow A_n})\label{eq:voltagedroppartition}\ \forall n\in \mathcal{N}_{\mathcal{G}}\setminus PCC(\mathcal{G}) \\
         &\boldsymbol{S}_{n\rightarrow A_n} = 
        \begin{cases}
            \boldsymbol{U}_n+\boldsymbol{u}_n \ \forall n\in \mathcal{M}_{\mathcal{G}}\\
            \sum_{m\in D_n} \boldsymbol{S}_{m\rightarrow n}\ \forall n\in\mathcal{N}_{\mathcal{G}}\setminus\mathcal{M}_{\mathcal{G}}
        \end{cases}\label{eq:nodalpowerbalancepartition}\\
        &\boldsymbol{\delta v}_{n=0} = 0\ \label{eq:refvoltagegridpartition}\\
        &|S_{n\rightarrow A_n}^t|^2\leq \overline{S}_{n\rightarrow A_n}^2 \ \forall n\in\mathcal{N}_{\mathcal{G}}\setminus PCC(\mathcal{G}), \forall t\in\mathcal{T} \label{eq:forwardpowerlimitpartition}\\
        &|\boldsymbol{\delta v}_n|\leq\overline{\delta v}_n \ \forall n\in\mathcal{N}_{\mathcal{G}}\setminus PCC(\mathcal{G})\label{eq:voltagelimitpartition}\\
        &g_n(\boldsymbol{u_n})\leq 0\ \forall n\in\mathcal{M}_{\mathcal{G}}\label{eq:feasibleflexpartition}
    \end{align}
\end{subequations}
Equations \eqref{eq:voltagedroppartition},\eqref{eq:nodalpowerbalancepartition} are voltage and power balance equations respectively. The PCC of the grid $PCC(\mathcal{G})$ is assumed to be maintained at the reference $v_{ref}$ at all times, in Eq \eqref{eq:refvoltagegridpartition}. The allowed magnitudes of branch power flows and voltage deviations are given in Eq \eqref{eq:forwardpowerlimitpartition},\eqref{eq:voltagelimitpartition}, and the feasibility region for flexibility dispatch, which is assumed to be convex for each prosumer, is given in  Eq \eqref{eq:feasibleflexpartition}. We denote the feasible region corresponding to constraints \eqref{eq:voltagedroppartition}-\eqref{eq:feasibleflexpartition} for the variables $\boldsymbol{u}\in\mathbb{C}^{|\mathcal{M}_{\mathcal{G}}|T}$ given the prosumptions $\boldsymbol{U}\in\mathbb{C}^{|\mathcal{M}_{\mathcal{G}}|T}$ as $\mathcal{B}_{\mathcal{G}}(\boldsymbol{U})$. 

Notice that from Eq \eqref{eq:voltagedroppartition}-\eqref{eq:refvoltagegridpartition}, voltage deviations and power flows are uniquely determined by linear combinations of the prosumptions $\boldsymbol{U}\in\mathbb{C}^{|\mathcal{M}_{\mathcal{G}}|T}$ and flexibility dispatches $\boldsymbol{u}$. We use the notation $\boldsymbol{S}(\boldsymbol{u},\boldsymbol{U};\mathcal{G})\in\mathbb{C}^{|\mathcal{E}_{\mathcal{G}}|T}$ to compactly denote the vector of branch power flows and $\boldsymbol{\delta v}(\boldsymbol{u},\boldsymbol{U};\mathcal{G})\in\mathbb{C}^{|\mathcal{N}_{\mathcal{G}}|T}$ to denote nodal voltage deviations (for all nodes and timesteps) given the flexibility dispatch $\boldsymbol{u}$ and inelastic prosumptions $\boldsymbol{U}$.


While the power flow model was formulated given the exact values prosumptions $\boldsymbol{U}$, in practice prosumptions are stochastic and one may only have access to their forecasts. In our work, we assume that point forecasts $\boldsymbol{\hat U}\in\mathbb{C}^{|\mathcal{M}_{\mathcal{G}}|T}$ are available for each prosumer in place of $\boldsymbol{U}$. A similar version of power flow equations \eqref{eq:voltagedroppartition}-\eqref{eq:refvoltagegridpartition} can then be written under the forecasts of prosumption $\boldsymbol{\hat U}$, and we denote the resulting feasible flexibility region using the notation $\mathcal{B}_{\mathcal{G}}(\boldsymbol{\hat U})$.

\section{Partition costs under two-stage energy dispatch}
\label{sec:probform}
We now formulate the two-stage cost model to calculate total cost under any given grid partition under uncertain prosumption. In this work, we focus on short-term energy dispatch costs for a given grid and prosumer configuration, while long term capacity investment decisions are not considered.

\begin{figure}[h!]
    \centering
    \includegraphics[width=0.9\linewidth]{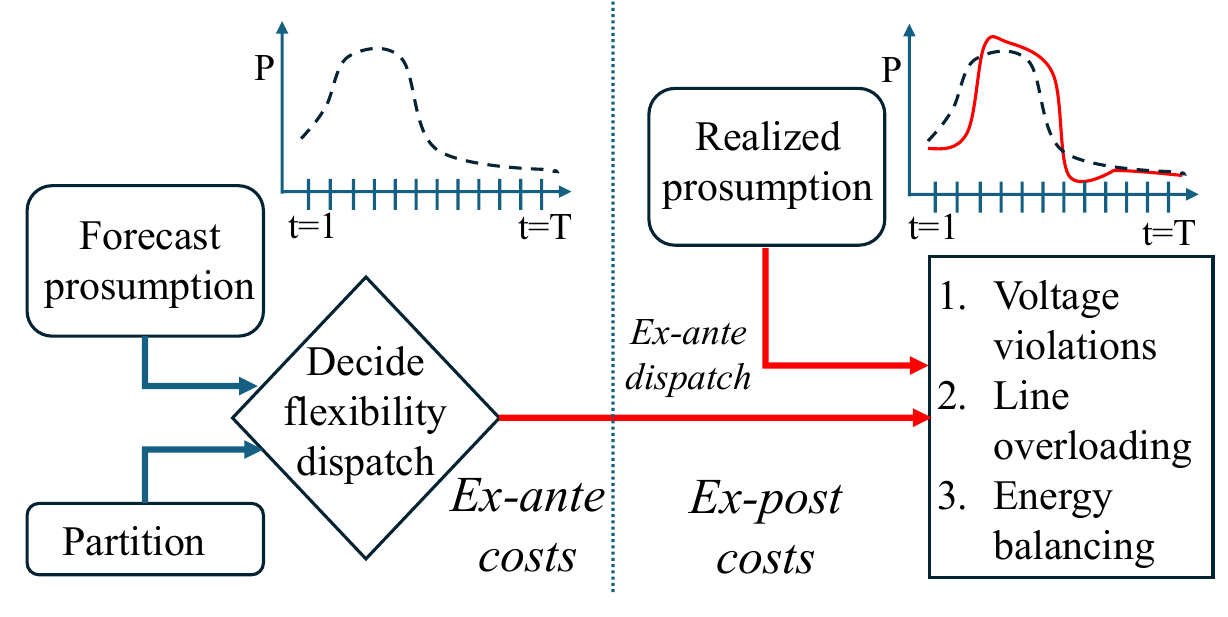}
    \caption{Two-stage costs under imperfect forecasts}
    \label{fig:twostagevisualise}
\end{figure}

In the first stage, given a grid partition, an optimal power flow problem is solved under forecasts of nodal prosumption, to determine the \emph{ex-ante} flexibility dispatch and costs. In the second stage, the uncertain prosumption is realized and \emph{ex-post} penalties are incurred for energy balancing and constraint violations. This is summarized in Fig \ref{fig:twostagevisualise}.

\subsection{Ex-ante stage}
\label{sec:exantesection}
In the \emph{ex-ante} stage, an optimal power flow problem is solved to minimize aggregate operating costs under the constraints Eq \eqref{eq:voltagedroppartition}-\eqref{eq:feasibleflexpartition} given forecasts of prosumption $\boldsymbol{\hat U}$. We let the aggregate costs consist of three terms: operating costs for flexible resources, costs for power imports from the transmission grid, and a tax penalty for each LEM's power exchanges with the rest of the distribution grid.

Denote the flexibility cost for dispatching $\boldsymbol{u}_n$ at prosumer node $n\in\mathcal{M}$ as $c_n(\boldsymbol{u}_n)$, with $c_n(\cdot)\in\mathbb{R}$ assumed to be convex which is standard in optimal power flow literature. The aggregate flexibility costs are then given as $\sum_{n\in\mathcal{M}}c_n(\boldsymbol{u}_n)$.

Denote the marginal price for power exchange at the PCC of the distribution grid ($n=0$) as the complex number $\lambda_0^t,\  t\in\mathcal{T}$, with its real and imaginary parts corresponding to active and reactive power price respectively. The total revenue for the grid from power exports, which are the sum of prosumer power injections, is given as $\left\langle\boldsymbol{\lambda}_0,\sum_{n\in\mathcal{M}_{\mathcal{G}}} (\boldsymbol{u}_n+\boldsymbol{\hat U}_n)\right\rangle$.

Finally, inspired by real-world tariff structures \cite{webmasterlausannech_electricite_2017}, we assume that any LEM coalition in the distribution grid faces a tax rate $\kappa_t>0$ on the magnitude of its power exchanges across its boundary nodes. For the LEM $F$, the total power exchanged at time $t$ and boundary node $n_b\in\mathcal{N}_F^b$ can be written as the following sum
\begin{align}
    &S^t_{n_b}(\boldsymbol{u},\boldsymbol{\hat U};F,\mathcal{G}) =  \sum_{(i\rightarrow n_b)\in\mathcal{E}_F} S^t_{i\rightarrow n_b}(\boldsymbol{u},\boldsymbol{\hat U};\mathcal{G})
    \label{eq:nodalexchangelem}
\end{align} 

As a result, for the partition $P=\{F_l\}_{l=1}^L$, the aggregate cost function for the optimal power flow problem under forecasts $\boldsymbol{\hat U}$ can be written as
\begin{equation}
\label{eq:firststagecosts}
    \begin{aligned}
        &\Gamma_e^{I}(\boldsymbol{u};P,\boldsymbol{\hat U}) = \sum_{l=1}^L \sum_{m\in\mathcal{M}_{F_l}}c_m (\boldsymbol{u}_m)-\left\langle\boldsymbol{\lambda}_0,\boldsymbol{\hat U}_m+\boldsymbol{u}_m\right\rangle \\
        &\qquad\qquad\quad+\sum_{l=1}^L\sum_{t\in\mathcal{T}}\sum_{n_b\in\mathcal{N}_{F_i}^b}\kappa_t \left\lVert S^t_{n_b}(\boldsymbol{u},\boldsymbol{\hat U};F_i,\mathcal{G}) \right\rVert_2
    \end{aligned}
\end{equation}

The \emph{ex-ante} dispatch, denoted as $\boldsymbol{u}^I(P,\boldsymbol{\hat U})$, can be written as the solution of the following optimal power flow problem
\begin{align}
    \boldsymbol{u}^I(P,\boldsymbol{\hat U})\in\argmin_{\boldsymbol{u}\in\mathcal{B}_{\mathcal{G}}(\boldsymbol{\hat U})} \Gamma_e^{I}(\boldsymbol{u};P,\boldsymbol{\hat U})
    \label{eq:firststagesolution}
\end{align}
The \emph{ex-ante} dispatch $\boldsymbol{u}^I(P,\boldsymbol{\hat U})\in\mathbb{C}^{|\mathcal{M}_{\mathcal{G}}|T}$ depends on the forecasts $\hat{\boldsymbol{U}}$ and the partition $P$ through the objective \eqref{eq:firststagecosts} and the constraints $\mathcal{B}_{\mathcal{G}}(\boldsymbol{\hat U})$. For brevity, we denote it using $\boldsymbol{u}^I$.

\subsection{Ex-post stage}
After the calculation of \emph{ex-ante} flexibility dispatch according to Eq \eqref{eq:firststagesolution}, the uncertain prosumptions $\boldsymbol{U}$ are realized. This results in the realized branch power flows $\boldsymbol{S}(\boldsymbol{u}^I,\boldsymbol{U};\mathcal{G})$ and realized nodal voltage deviations $\boldsymbol{\delta v}(\boldsymbol{u}^I,\boldsymbol{U};\mathcal{G})$, which may be different from their anticipated \emph{ex-ante} values. This may result in violation of constraints \eqref{eq:forwardpowerlimitpartition},\eqref{eq:voltagelimitpartition}. 

We assume that the cost of constraint violations is proportional to the magnitude of the violation, with $\alpha_{e}^{\Delta S}, \alpha_n^{\Delta v}$ being the cost sensitivities for line limit violations in edge $e$ and voltage violations in node $n$ respectively. The \emph{ex-post} costs for voltage and line violations are written as
\begin{subequations}
\label{eq:voltageandlinecostspartition}
\begin{align}
        \begin{split}
        &\Gamma_c^{II}(\boldsymbol{u}^I,\boldsymbol{U}) = \sum_{t=1}^T\left(\Gamma_{c,0}^t(\boldsymbol{u}^I,\boldsymbol{U})+\sum_{l=1}^{L}\Gamma^t_{c,F_l}(\boldsymbol{u}^I,\boldsymbol{U})\right)
        \label{eq:totalcostsconstraintviolationsbreakup}
        \end{split}
        \\
        &\mathrm{where}\nonumber \\
        \begin{split}
        &\Gamma^t_{c,F_l}(\boldsymbol{u}^I,\boldsymbol{U}) = \sum_{e\in\mathcal{E}_{F_i}}\alpha_{e}^{\Delta S}(\lVert \boldsymbol{S}_{e}^{t}(\boldsymbol{u}^I,\boldsymbol{U};\mathcal{G})\rVert_2-\overline{S}_{e})^+\\
        &\qquad \qquad\quad+\sum_{n\in\mathcal{N}_{F_l}^{int}}\alpha_n^{\Delta v}(|\boldsymbol{\delta v}_n^{t}(\boldsymbol{u}^I,\boldsymbol{U};\mathcal{G})|-\overline{\delta v}_n)^+
        \label{eq:lemconstraintviolations}
        \end{split}\\
        \begin{split}
        &\Gamma^t_{c,0}(\boldsymbol{u}^I,\boldsymbol{U})= \sum\limits_{e\in\mathcal{E}_{ext}}\alpha_{e}^{\Delta S}(\lVert \boldsymbol{S}_{e}^{t}(\boldsymbol{u}^I,\boldsymbol{U};\mathcal{G})\rVert_2-\overline{S}_{e})^+\label{eq:commonconstraintviolations}\\
        &\quad\qquad\qquad+\sum\limits_{n\in\mathcal{N}_{ext}}\alpha_n^{\Delta v}(|\boldsymbol{\delta v}_n^{t}(\boldsymbol{u}^I,\boldsymbol{U};\mathcal{G})|-\overline{\delta v}_n)^+
        \end{split}
\end{align}
\end{subequations}
Here Eq \eqref{eq:lemconstraintviolations} is the cost for constraint violations within LEM $F_l$'s subgraph. Eq \eqref{eq:commonconstraintviolations} is the cost for constraint violations in the nodes $\mathcal{N}_{ext}=\mathcal{N}_{\mathcal{G}}\setminus\cup_{l=1}^L \mathcal{N}_{F_l}^{int}$ and edges $\mathcal{E}_{ext}=\mathcal{E}_{\mathcal{G}}\setminus\cup_{l=1}^L \mathcal{E}_{F_l}$, which are external to all the LEMs. Note that the total cost of constraint violations (Eq \eqref{eq:totalcostsconstraintviolationsbreakup}) depends on the partition $P$ and forecasts $\boldsymbol{\hat U}$ via the \emph{ex-ante} dispatch $\boldsymbol{u}^I$.

In addition to constraint violations, imperfect forecasts may result in energy imbalances in the grid. We assume that the penalty for imbalance for the DSO is proportional to the magnitude of the total energy imbalance at rate $\alpha_{0}^{\Delta E}$. 

The total balancing costs are then given by\footnote{It may be possible to jointly mitigate constraint violations and energy imbalance, but in this work we use a conservative estimate for simplicity.}
\begin{align}
\begin{aligned}
    &\Gamma_{b}^{II}(\boldsymbol{\hat U},\boldsymbol{U}) = \sum_{t=1}^T\Gamma_b^t (\boldsymbol{\hat U},\boldsymbol{U})\\
    &\mathrm{where\ }\Gamma_b^t (\boldsymbol{\hat U},\boldsymbol{U}) = \alpha_{0}^{\Delta E} \left\lVert\sum_{m\in \mathcal{M}}\left(U_m^t-\hat U_m^t\right)\right\rVert_2
\end{aligned}
\label{eq:energybalancingpartition}
\end{align}
\subsection{Total costs for a grid partition}
Using the costs and dispatch calculated in \emph{ex-ante} and \emph{ex-post} stages, the total costs for a partition $P$ are written as
\begin{align}
\fbox{$
\begin{aligned}
&\Phi(P;\boldsymbol{\hat U},\boldsymbol{U})
=  \Gamma_c^{II}(\boldsymbol{u}^I(P,\boldsymbol{\hat U}),\boldsymbol{U})+\Gamma_{b}^{II}(\boldsymbol{\hat U},\boldsymbol{U})\\&\qquad\qquad\qquad +\Gamma_e^{I}(\boldsymbol{u}^I(P,\boldsymbol{\hat U});P,\boldsymbol{U})
\label{eq:totalcostspartition}
\end{aligned}
$}
\end{align}
These are the total costs for the entire distribution grid, and their minimization is the objective of the DSO. For brevity, we shall use $\Phi(P)$ to denote $\Phi(P;\boldsymbol{\hat U},\boldsymbol{U})$. 
\section{Coalition cost model}
\label{sec:coalitioncostssection}
In the previous section, we discussed the two stage cost for a grid partition (Eq \eqref{eq:totalcostspartition}). This defines the total cost incurred for a grid partition, which consists of multiple LEM coalitions. To study the partitioning problem from prosumers' perspective, we need to define a model for cost incurred by an LEM coalition in the distribution grid.

\begin{figure}[h!]
    \centering
    \includegraphics[width=0.9\linewidth]{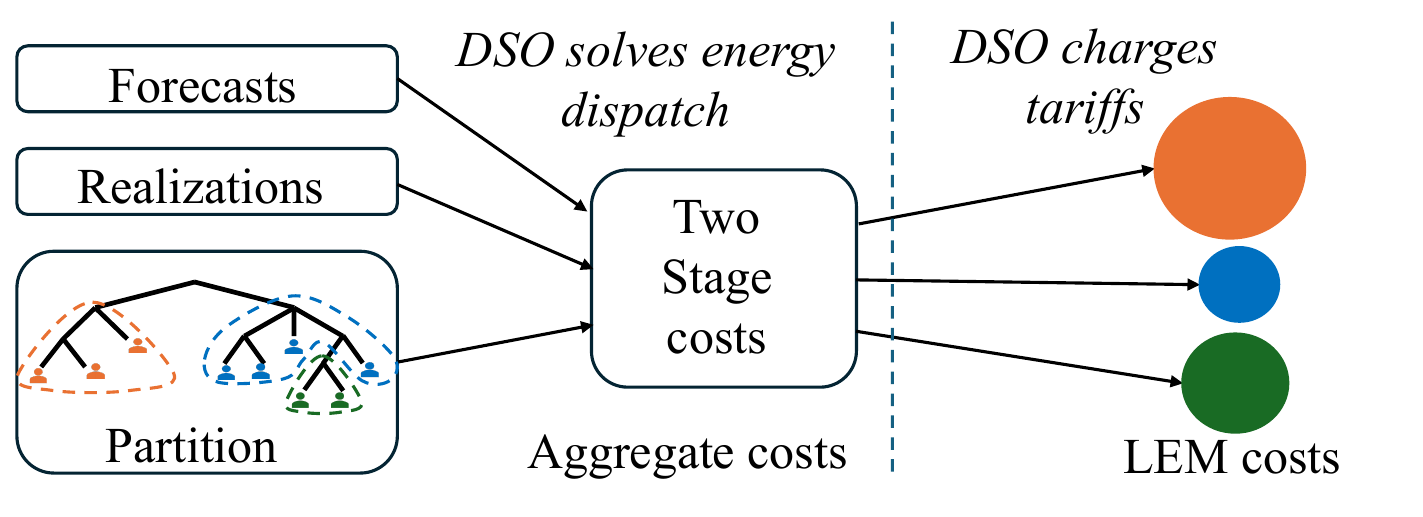}
    \caption{Cost recovery via tariffs to LEMs in a partition}
    \label{fig:tariffvisualize}
\end{figure}

We assume that the DSO is revenue-neutral, and sets an energy tariff rule to LEMs to recover its total operating costs (Eq \eqref{eq:totalcostspartition}). This induces a model for the coalition costs (Fig \ref{fig:tariffvisualize}), which we now discuss.

\subsection{Coalition cost under coupled power flow constraints}
The total grid costs in Eq \eqref{eq:totalcostspartition} consist of two parts: energy costs, which comprise of imports, flexibility and taxes (Eq \eqref{eq:firststagecosts}); and costs incurred due to uncertainty, which comprise of constraint violation costs (Eq \eqref{eq:totalcostsconstraintviolationsbreakup}) and imbalance costs (Eq \eqref{eq:energybalancingpartition}). To recover these costs we assume that the DSO treats LEMs as virtual power plants, and charges a two part tariff for energy imports and imbalances. 

For the first part, we assume that each LEM is charged for power flows through its boundary nodes using the well-known distribution locational marginal price \cite{papavasiliou_analysis_2018}, derived using the dual variables for constraints \eqref{eq:nodalpowerbalancepartition} in the problem Eq \eqref{eq:firststagecosts}. 

For the second part of the tariff, we assume that the DSO charges an imbalance penalty to each LEM depending on the magnitude of its net prosumption imbalance, in order to recover the shared \emph{ex-post} costs for energy balancing Eq \eqref{eq:energybalancingpartition} and constraint violations in common nodes and edges  Eq \eqref{eq:commonconstraintviolations}.


The above define the costs charged by the DSO to an LEM for its interactions with the grid. Apart from these, each LEM incurs flexibility costs, tax for power exchanges at rate $\kappa_t$ (both described in Section \ref{sec:exantesection}), and discomfort costs for voltage and line limit violations within its internal network (Eq \eqref{eq:lemconstraintviolations}). 

The total costs for coalition $F_i$ in partition $P$ are written as
\begin{subequations}
\label{eq:coalitioncostsexternalities}
\begin{align}
    \begin{split}
    &\phi(F_i;P)=\phi_{int}(F_i;P)+\phi_{ext}(F_i;P)
    \end{split}\\
    \begin{split}
       &\phi_{int}(F_i;P) = \sum_{n\in\mathcal{M}_{F_i}}c_n(\boldsymbol{u}_n^I)+\sum_{t\in\mathcal{T}}\Gamma_{c,F_i}^t(\boldsymbol{u}^{I},\boldsymbol{U})\\
       &\qquad\qquad\qquad+\sum_{t\in\mathcal{T}}\sum_{n_b\in\mathcal{N}_{F_i}^b}\kappa_t \left\lVert S^t_{n_b}(\boldsymbol{u}^{I},\boldsymbol{U};F_i) \right\rVert_2
    \label{eq:intcostscoalition}
    \end{split}
    \\
    \begin{split}
       &\phi_{ext}(F_i;P) = -\sum_{n_b\in\mathcal{N}_F^b}\left\langle\boldsymbol{\lambda}_{n_b},\boldsymbol{S}_{n_b}(\boldsymbol{u}^{I},\boldsymbol{U};F_i)\right\rangle\\
       &\qquad\qquad\qquad+\sum_{t\in\mathcal{T}}\alpha_t^{\Delta E}\left\lVert\sum_{n\in\mathcal{M}_F}\hat U_n^t-U_n^t\right\rVert_2
    \label{eq:extcostscoalition}
    \end{split}
    \\
    \begin{split}
       &\mathrm{where}\quad  \boldsymbol{\lambda}_n \ \mathrm{is\ dual\ variable\ for\ } \eqref{eq:nodalpowerbalancepartition} \mathrm{\ in\ } \eqref{eq:firststagecosts}, 
    \end{split}
    \\
    \begin{split}
        &\qquad\qquad \alpha_t^{\Delta E} = \frac{\Gamma_{c,0}^t(\boldsymbol{u}^I,\boldsymbol{U})+\Gamma_b^t(\boldsymbol{\hat U},\boldsymbol{U})}{\sum_{l=1}^L \left\lVert\sum_{n\in\mathcal{M}_{F_l}}\hat U_n^t-U_n^t \right\rVert_2}
    \label{eq:balancingcoeffexpost}
    \end{split}
\end{align}
\end{subequations}

Eq \eqref{eq:intcostscoalition} refers to costs faced by the coalitions in its internal network and Eq \eqref{eq:extcostscoalition} refers to the costs as a result of its interactions with the distribution grid. Eq \eqref{eq:balancingcoeffexpost} sets the imbalance penalty for cost recovery for \emph{ex-post} costs in common nodes/edges \eqref{eq:energybalancingpartition}, \eqref{eq:commonconstraintviolations}. Both Eq \eqref{eq:totalcostspartition}, \eqref{eq:coalitioncostsexternalities} include the effect of uncertainty and network constraint violations, which was not addressed in previous works.
 
\textbf{Remark:} Observe that the total cost $\phi(F_i;P)$ for a coalition $F_i$ depends on the grid partition $P$, and hence on the partitioning of prosumers external to the coalition through the prices $\boldsymbol{\lambda}, \boldsymbol{\alpha}^{\Delta E}$ and the dispatches $\boldsymbol{u}^I$. These happen due to coupled power flow constraints Eq \eqref{eq:voltagedroppartition}-\eqref{eq:voltagelimitpartition}, and are termed as \emph{externalities}, with an example illustrated in Fig \ref{fig:costexternalities}\footnote{See Appendix \ref{app:simulationslausanne} for a numerical example on a real grid}. 

\begin{figure}[h!]
    \centering
    \includegraphics[width=0.95\linewidth]{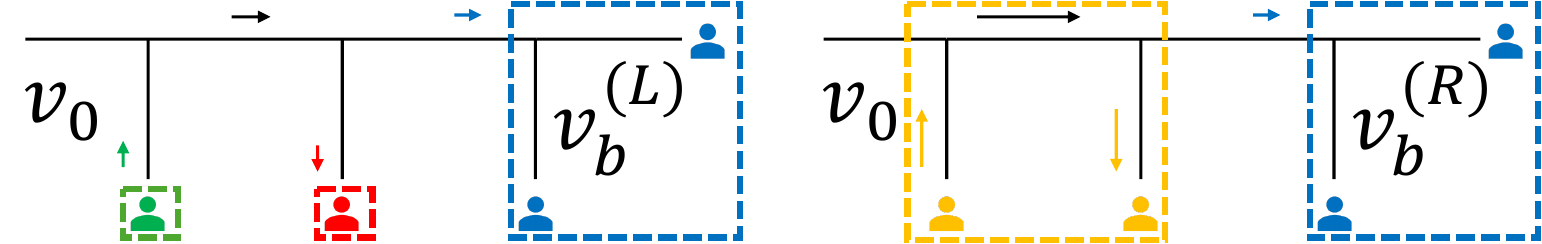}
    \caption{An example of externalities. The voltage at the PCC for the blue LEM coalition depends on upstream LEM configuration. If upstream prosumers self-consume in separate coalitions (left), there is reduced flow and voltage drop across their connecting edge compared to forming single LEM (right). In congested grids with uncertainty, the blue LEM's ex-post costs can thereby be influenced by upstream prosumers' configuration.}
    \label{fig:costexternalities}
\end{figure}
Analyzing partition stability becomes complicated under externalities, since coalitional deviations must account for configurations of the remainder grid to analyze coalition costs under deviation, as will be discussed later.
\subsection{Coalition cost under decoupled power flow}
While in general the costs incurred by each LEM \eqref{eq:coalitioncostsexternalities} may depend on the partition, under a special case the costs of a coalition can be written independently of the partitioning of the remainder grid. This requires the following assumptions.
\begin{assumption}
    (LEMs with single boundary node):
    Prosumers are allowed to form any LEM $F\in\mathcal{S}(\mathcal{G})$ only if $|\mathcal{N}_F^b|=1$, that is, it has exactly one boundary node $PCC(F)$
    \label{assm:boundarynode}
\end{assumption}
Currently, local energy markets typically have a single net-metering point, so Assumption \ref{assm:boundarynode} is realistic.
\begin{assumption}
    (Strict self consumption): The DSO sets $\kappa_t\rightarrow\infty$, and all LEMs satisfying Assumption \ref{assm:boundarynode} ensure zero energy exchange with the grid in \emph{ex-ante} and \emph{ex-post} stages, that is $\boldsymbol{S}_{PCC(F)}(\boldsymbol{u},\boldsymbol{U};F) = \boldsymbol{S}_{PCC(F)}(\boldsymbol{u},\boldsymbol{\hat U};F) = 0$
    \label{assm:strictselfconsume}
\end{assumption}
Under Assumption \ref{assm:strictselfconsume}, each LEM must independently ensure energy balancing, and we let $\alpha_{F_i}^{\Delta E}$ denote its marginal price for energy balancing. While Assumption \ref{assm:strictselfconsume} represents the ideal case of self-consumption and may not be realistic, it provides insights on the effect of uncertainty on coalition formation, which we will discuss in later sections. In realistic grids, this assumption may not always hold. Then, one would use the coalition cost model for externalities (Eq \eqref{eq:coalitioncostsexternalities}) to analyze coalition formation, as we discuss in Section \ref{sec:optimalstablepartition}.

Under these assumptions, the coalition costs Eq \eqref{eq:coalitioncostsexternalities} can be rewritten using the following Proposition.
\begin{proposition}
\label{prop:costscoalitionnoexternalities}
    (Coalition costs for strict self-consumption and single boundary node) For any $P=\{F_i\}_{i=1}^L$ under Assumptions \ref{assm:boundarynode}-\ref{assm:strictselfconsume}, the coalition costs \eqref{eq:coalitioncostsexternalities} are given as
    \begin{subequations}
    \label{eq:coalitioncostsnoexternalities}
        \begin{align}
            &\phi(F_i,P) := \phi_0(F_i) = \Gamma_{e0}^I(F_i)+\Gamma_{c0}^{II}(F_i)+\Gamma_{b0}^{II}(F_i)\\
            &\Gamma_{e0}^I(F_i) = \sum_{n\in\mathcal{M}_{F_i}}c_n((\boldsymbol{u}_{F_i}^I)_n)\label{eq:coalitioncostsexanteproposition}\\
            &\Gamma_{b0}^{II}(F_i) = \sum_{t\in\mathcal{T}}\alpha_{F_i}^{\Delta E}\left\lVert\sum_{m\in \mathcal{M}_{F_i}}\left(U_m^t-\hat U_m^t\right)\right\rVert_2\label{eq:balancecostscoalitionsproposition}\\
            \begin{split}
            &\Gamma_{c0}^{II}(F_i) = \sum_{t=1}^T\sum_{e\in\mathcal{E}_{F_i}}\alpha_{e}^{\Delta S}(\|\boldsymbol{S}_e^{t}(\boldsymbol{u}_{F_i},\boldsymbol{U};F_i)\|_2-\overline{S}_{e})^+\\
            &\qquad\quad+\sum_{t=1}^T\sum_{n\in\mathcal{N}_{F_i}^{int}}\alpha_n^{\Delta v}(|\boldsymbol{\delta v}_n^{t}(\boldsymbol{u}_{F_i},\boldsymbol{U};F_i)|-\overline{\delta v}_n)^+
            \label{eq:coalitioncostsinternalproposition}
            \end{split}\\
            &\mathrm{where}\ \boldsymbol{u}_{F_i}^I\in\argmin_{\boldsymbol{u}_{F_i}\in\mathcal{B}_{F_i}(\boldsymbol{\hat U})}\sum_{n\in\mathcal{M}_{F_i}}c_n\left((\boldsymbol{u}_{F_i})_n\right)\label{eq:exanteobjectivecoalitionproposition}\\
            &\qquad\qquad \quad\mathrm{s.t.}\  \boldsymbol{S}_{PCC(F_i)}(\boldsymbol{u},\boldsymbol{\hat U};F_i) =0\label{eq:boundaryzeroexchangecoalitionproposition}
        \end{align}
    \end{subequations}
    Here $\mathcal{B}_{F_i}(\boldsymbol{\hat U})$, \eqref{eq:boundaryzeroexchangecoalitionproposition} are power flow constraints for radial graph for coalition $F_i$, and $\boldsymbol{S}(\boldsymbol{u}_{F_i},\boldsymbol{U};F_i), \ \boldsymbol{\delta v}(\boldsymbol{u}_{F_i},\boldsymbol{U};F_i)$ are branch power flows and nodal voltage deviation functions corresponding to these power flow equations.
\end{proposition}

The proof is given in the Appendix \ref{app:prop1}. In this case, the coalition cost function $\phi(F_i;P)$ is given as $\phi_0(F_i)$, and only depends on the coalition $F_i$ but not the partition $P$. This corresponds to partitioning with \emph{no externalities}, where the costs of a coalition do not depend on the remainder grid. This is a special case\footnote{Although Eq \eqref{eq:coalitioncostsnoexternalities} is a special case of Eq \eqref{eq:coalitioncostsexternalities}, which is under a specific pricing model, it could also be independently derived by decomposing \eqref{eq:totalcostspartition} using Assumptions \ref{assm:boundarynode}-\ref{assm:strictselfconsume}} of Eq \eqref{eq:coalitioncostsexternalities}. This simplifies the analysis of partition stability, because coalitional deviations can be analyzed independently of the remainder grid, as we discuss in later sections. 

We analyze both coalition cost models (with and without externalities) for stable partitioning in Section \ref{sec:optimalstablepartition}.

\section{Partitioning objectives}
We now formulate the partitioning objectives from the DSO's and the prosumers' perspective.
\label{sec:partitioningoutcomes}
\subsection{DSO's perspective: optimal partitioning}
\label{sec:partitioningoutcomesdso}
The problem of finding the optimal partition, \textbf{from the DSO's perspective}, refers to calculating the partition that minimizes the aggregate costs given in Eq \eqref{eq:totalcostspartition}. The optimal partition $P_{DSO}^{\star}$ can be written as the solution of the following
\begin{align}
\fbox{$
\begin{aligned}
P_{DSO}^{\star} & \in \argmin_{P\in\mathcal{P}(\mathcal{G})}\Phi(P)
\end{aligned}
$}
\tag{I}
\label{eq:optimalpartitiondso}
\end{align}
Problem \eqref{eq:optimalpartitiondso} is an optimization problem over the set of partitions $\mathcal{P}(\mathcal{G})$ of the graph $\mathcal{G}$, and determining the optimal partition in general would require iterating over $\mathcal{P}(\mathcal{G})$. Since the number of partitions of a set scale super-exponentially, the optimization problem is \textsc{np}-hard.

Under perfect forecasts, however, it is easy to find the solution for Problem \eqref{eq:optimalpartitiondso}, as follows.

\begin{proposition}
    (Optimal partition under perfect forecasts) Let $P_{GC}:=\{(\mathcal{N},\mathcal{E})\}$ denote the partition corresponding to the largest LEM in the distribution grid. For $\boldsymbol{U}=\boldsymbol{\hat U}$, $P_{GC}$ is the minimizer solution for Problem \eqref{eq:optimalpartitiondso}
    \label{prop:optimalityperfectforecasts}
\end{proposition}
The proof is given in Appendix \ref{app:lemma1} and uses the fact that the largest LEM minimizes the \emph{ex-ante} costs \eqref{eq:firststagecosts}, and that there are no \emph{ex-post} costs under perfect forecasts. 

In contrast to the above theorem, for the setting of imperfect forecasts, the optimal partition may not correspond to $P_{GC}$, and partitioning into smaller LEMs may be beneficial. 

To understand this, note that voltage deviations \eqref{eq:voltagedroppartition} and line power flows \eqref{eq:refvoltagegridpartition} are linear functions of prosumers' injections. Increased trading in larger LEMs may cause higher power flows and voltage deviations, increasing the likelihood of constraint violations and incurring higher \emph{ex-post} costs under uncertain prosumption. This is illustrated in Example \ref{example:1} below.
\begin{example}
\label{example:1}
    Consider a simple radial grid consisting of one root node and three leaf nodes, with prosumers $1, 2, 3$ located at its leaf nodes. Let the time horizon be $T=3$, and the prosumption forecasts and realizations be given in Table \ref{tab:ex1prosumdata}.
\begin{table}[h!]
    \centering
    \begin{tabularx}{\linewidth}{|Y|Y|Y|Y|Y|Y|Y|}
    \hline
    \multirow{2}{*}{} & \multicolumn{3}{c|}{Forecast} & \multicolumn{3}{c|}{Realized}\\
    \cline{2-7}
     & $t=1$&$t=2$&$t=3$&$t=1$&$t=2$&$t=3$\\
    \hline
    P1 & 0 & 1 & -1 & 0 &1.2&-1.2\\
    P2 & -1 & 0 & 1&-1.2&0 & 1.2 \\
    P3 & 1 & -1 & 0 & 1.2 &-1.2&0\\
    \hline
    \end{tabularx}
    \caption{Forecasts and realizations for Example 1}
    \label{tab:ex1prosumdata}
\end{table}

Let each prosumer have a battery for flexibility, with identical operating costs $19\ CHF/MWh$. Each prosumer can either operate its battery for energy storage at this cost, or trade energy in LEMs. Let $\kappa_t = 100 \ CHF/MWh, \lambda_0^t = 0$ and penalties for line overloading and energy imbalance be equal to $200\  CHF/MWh$. Let each of the lines that connect prosumers to the root node have limits of $1 \ MWh$, and ignore voltage constraints for now.

While there are five possible partitions for the grid, due to participant symmetry over the time horizon, we only need to analyze three partitions. To determine the flexibility dispatch, note that the tax rate $\kappa_t$ is more than battery costs, so any LEM will prefer to operate batteries to ensure net zero prosumption in \emph{ex-ante} stage instead of imports/exports from the external grid. Realized power flows, ex-post costs can be calculated following discussion in preceding sections. The costs for each of the partitions are summarized in Table \ref{tab:ex1costs}.

\begin{table}[h!]
    \centering
    \begin{tabular}{|c|c|c|c|c|c|}
        \hline
        \multirow{2}{*}{Partition} & \multicolumn{5}{c|}{Costs} \\
        \cline{2-6}
        & Flex & Imb & Over & Tax & Total \\
        \hline
        $\{\{1,2,3\}\}$ & $0$ & $0$ & $240$ & $0$ & $240$ \\
        \hline
        $\{\{1,2\},\{3\}\}$ & $76$ & $0$ & $80$ & $80$ & $236$ \\
        \hline
        $\mathbf{\{\{1\},\{2\},\{3\}\}}$ & $114$ & $0$ & $0$ & $120$ & $\mathbf{234}$ \\
        \hline
    \end{tabular}
    \caption{Costs under forecast errors for Example 1. Flex.: flexibility costs (\emph{ex-ante}); 
    Imb.: imbalance costs (\emph{ex-post}); 
    Over: Overload costs (\emph{ex-post}); 
    Tax: Tax payments (\emph{ex-post})}
    \label{tab:ex1costs}
\end{table}

Notice that while ex-ante flexibility costs are minimum in partition $\{\{1,2,3\}\}$, this partition results in very high costs for line overloading. Compared to this $\{\{1\},\{2\},\{3\}\}$ provides lower total costs by self-consuming to minimize constraint violations and is the optimal partition for the DSO.

\end{example}

Example \ref{example:1} shows that with imperfect forecasts, the DSO may prefer grid partitioning to reduce constraint violations. In general, whether partitioning is beneficial or not depends on the grid configuration, costs and forecast error; and one must solve Problem \eqref{eq:optimalpartitiondso} to determine the optimal partition.
\subsection{Prosumers' perspective: stable partitioning}
\label{sec:stablepartitioningsection}
While the DSO's objective is to find the partition to minimize the costs $\Phi(P)$ (Problem \eqref{eq:optimalpartitiondso}), the prosumers that form any LEM $F_i\in P$ face the costs $\phi(F_i;P)$ (Eq \eqref{eq:coalitioncostsexternalities}). Self-interested prosumers are thus likely to organize into coalitions that ensure minimum costs for themselves, which may not be aligned with the DSO's objective.


Once a partition $P$ is selected, each of its constituent LEM coalitions $F_i\in P$ must allocate their experienced costs $\phi(F_i,P)$ (Eq \eqref{eq:coalitioncostsexternalities}) amongst their constituent prosumers $j\in \mathcal{M}_{F_i})$. If, however, for any LEM coalition $F_i\in P$, some subset of its constituent prosumers find that they are allocated higher costs compared to their costs of forming alternative LEMs $\tilde F_i$, they may break off from this coalition. This can result in the formation of alternative coalitions, and thus alternative grid partitions, as illustrated in Example \ref{example:2}.
\begin{example}
\label{example:2}
Consider a modification of Example \ref{example:1}, where battery costs are modified to $19 \ CHF/MWh+2\ CHF/MWh^2$ from $19 \ CHF/MWh$ previously. Similar to Example \ref{example:1}, due to participant symmetry, we only analyze three partitions and three coalitions, whose costs are given in Tables \ref{tab:ex2partitioncosts}, \ref{tab:ex2coalitioncosts}.

{ 
\newcounter{SavedFigCount}
\setcounter{SavedFigCount}{\value{figure}} 

\setcounter{figure}{2}
\setcounter{table}{3}
\renewcommand{\thefigure}{\Roman{figure}}
\makeatletter
\renewcommand{\figurename}{TABLE}
\renewcommand{\thesubfigure}{(\Roman{figure}~\alph{subfigure})}
\renewcommand{\p@subfigure}{} 
\makeatother

\begin{figure}[h!]
\centering
\begin{subfigure}{0.48\linewidth}
    \centering
    \begin{tabular}{|c|c|}
    \hline
         Partition & Costs \\ \hline
         $\{\{1,2,3\}\}$ & 240 \\ \hline
         $\{\{1,2\},\{3\}\}$ & 241 \\ \hline
         $\{\{1\},\{2\},\{3\}\}$ & 252 \\ \hline
    \end{tabular}
    \\ \textbf{(a)} 
    \phantomcaption 
    \label{tab:ex2partitioncosts}
\end{subfigure}
\hfill
\begin{subfigure}{0.48\linewidth}
    \centering
    \begin{tabular}{|c|c|}
    \hline
         Coalition & Costs \\ \hline
         $\{1,2,3\}$ & 240 \\ \hline
         $\{1,2\}$ & 157 \\ \hline
         $\{3\}$ & 84 \\ \hline
    \end{tabular}
    \\ \textbf{(b)} 
    \phantomcaption
    \label{tab:ex2coalitioncosts}
\end{subfigure}
 
\caption{Example 2: (a) Partition costs, (b) Coalition costs.}
\end{figure}

\setcounter{figure}{\value{SavedFigCount}} 
} 
\vspace{-8pt}

In this example, the partition $\{\{1,2,3\}\}$ representing a single LEM coalition yields the lowest total costs $240 \ CHF$. Thus the \textbf{DSO will prefer} this partition. 

If this partition is formed, the DSO will collect $80 \ CHF$ from each prosumer equally (due to symmetry) for cost recovery. As a result, prosumers $1$ and $2$ will be charged a total of $160\ CHF$. However, note that if they instead form a separate LEM $\{1,2\}$, they would obtain costs equal to $157\  CHF$, which are lower than their cost allocation in the partition $\{\{1,2,3\}\}$. Thus, they would break off from the partition preferred by the DSO in favour of the partition $\{\{1,2\},\{3\}\}$.
\end{example}

Example \ref{example:2} indicates that it is not sufficient to select the partition that minimizes the aggregate costs (Problem \eqref{eq:optimalpartitiondso}). One must also study cost sharing to analyze whether a partition is stable with regards to deviations by any group of prosumers. We now formalize cost sharing for stable partitioning. We first define the \emph{core} \cite{shapley_game_1973} as follows

\begin{definition}
    (Core for a coalition $F_i$ in partition $P$): Given a LEM coalition $F_i\in P$, $Core(F_i;P)$ is defined as the set of cost allocations as follows
    \begin{align*}
        &Core(F_i;P)
        = \{\boldsymbol{y}_{F_i}\in\mathbb{R}^{|\mathcal{E}_{F_i}|}:  \sum_{e\in \mathcal{E}_{F_i}}(\boldsymbol{y}_{F_i})_e = \phi(F_i;P), \\
        &\qquad\qquad\qquad\quad\sum_{e\in \mathcal{E}_F} (\boldsymbol{y}_{F_i})_e\leq \hat \phi(F)\ \forall F\in \mathcal{S}(F_i)\}
    \end{align*}
    \label{def:coregeneral}
\end{definition}
Here, the equality constraint requires strict budget balance for each coalition $F_i\in P$, that is the cost allocation $\boldsymbol{y}_{F_i}$ over the graph $F_i$ must sum to the coalition cost $\phi(F_i;P)$, as discussed in Section \ref{sec:coalitioncostssection}. The inequality constraint requires that any prospective deviating coalition $F$ should not obtain higher allocated costs compared to its deviation costs $\hat \phi(F)$. 

A non-empty core implies that there exists a cost allocation $\boldsymbol{y}_{F_i}$ that ensures no subset of prosumers prefer to split off from the coalition $F_i$ to form an alternative coalition. We call a coalition stable, if its \emph{core} is nonempty. 

To calculate the core, one needs the deviation costs $\hat \phi(F)$ for any coalition $F$. This can be complicated due to externalities, where coalition costs may depend on the remainder grid due to coupled power flow constraints, as discussed in Section \ref{sec:coalitioncostssection}. Hence, each deviating coalition $F$ must anticipate the partitioning of $\mathcal{G}\setminus F$ following its deviation to calculate $\hat \phi(F)$. 

We now provide a model for the estimates $\hat \phi(F)$ under two cases: Partitioning under \emph{no externalities}, and partitioning with \emph{externalities} for coalition costs discussed in Section \ref{sec:coalitioncostssection}. 
\begin{case}
    \emph{Deviation costs under no externalities}: In a special case, when coalitions satisfy Assumptions \ref{assm:boundarynode}-\ref{assm:strictselfconsume}, the coalition cost can be written using Eq \eqref{eq:coalitioncostsnoexternalities}. In this case, the costs for any coalition $F$ only depend on its internal configuration, and not on how the remainder grid $\mathcal{G}\setminus F$ is partitioned. Hence, deviation cost $\hat \phi(F)$ can be written as $\phi_0(F)$ from Eq \eqref{eq:coalitioncostsnoexternalities}, and coalition stability can be studied using the characteristic function formulation \cite{shapley_game_1973}.

    This implies that a LEMs coalition's decision to split into smaller LEMs only depends on its internal energy trading and the prosumption uncertainty. This case allows us to isolate the effect of uncertainty on formation of LEM partitions.
    \label{case:noexter}
\end{case}

\begin{case}
\emph{Deviation costs under externalities: } In this case, coalition costs are given by Eq \eqref{eq:coalitioncostsexternalities}, and depend on the partitioning of the remainder grid. Hence, to calculate $\hat{\phi}(F)$ for any deviating coalition $F$, one must model how the coalition anticipates the remainder grid $\mathcal{G}\setminus F$ to react to its deviation. This is studied formally in partition function form games \cite{thrall_nperson_1963}. 


In this work, following the model provided by the authors \cite{hart_endogenous_1983} in partition function form games, we assume that the remainder prosumers remain in their existing coalitions and no new cooperative links are formed, reflecting inertia in reactions of prosumers.

As a result, when coalition $F$ deviates from $F_i$ in the partition $P=\{F_i\}_{i=1}^L$, the resulting partition is $P'=\{F,F_i\setminus F, F_j\ \mathrm{for\ }j\neq i\}$. The deviation costs $\hat \phi(F)$ given the partition $P$ are then given as $\hat \phi(F) = \phi(F;P')$.
\label{case:exter}
\end{case}

\medskip

Having now examined deviation costs $\hat \phi(F)$ in the definition of the core for a coalition (Def \ref{def:coregeneral}), we now define a stable partition, adapted from \cite{abada_unintended_2020}, as 
\begin{definition}
    (Stable partition): A partition $P=\{F_i\}_{i=1}^L$ is denoted as a stable if and only if $Core(F_i;P)\neq \emptyset\ \forall F_i\in P$
\label{def:stablepartition}
\end{definition}
A stable partition implies that self-interested prosumers have no incentive to split from their current LEM coalitions, and is the desired outcome from the \textbf{prosumers' perspective}. As Example \ref{example:2} shows, the stable partition may not coincide with the optimal partition (\textbf{DSO's perspective}).

Under the above definition, one can verify that the existence of a stable partition is ensured since the partition comprised of singleton coalitions is always stable.

\section{Optimal stable partitioning}
\label{sec:optimalstablepartition}
To balance the DSO's interests, who wants to minimize the aggregate cost \eqref{eq:totalcostspartition}, and that of prosumers, who prefer stable partitions, we formulate the optimal stable partition as
\begin{align}
\fbox{$
    \begin{aligned}
        &\Gamma^{\star} = \min_{P\in \mathcal{P}(\mathcal{G})} \Phi(P)\\
        &\qquad\qquad\mathrm{s.t.\ } P \ \mathrm{satisfies\ Def\ } \ref{def:stablepartition}
    \end{aligned}
$}
\tag{II}
\label{eq:optimalstablepartition}
\end{align}

\begin{figure}[h!]
    \includegraphics[width=0.9\linewidth]{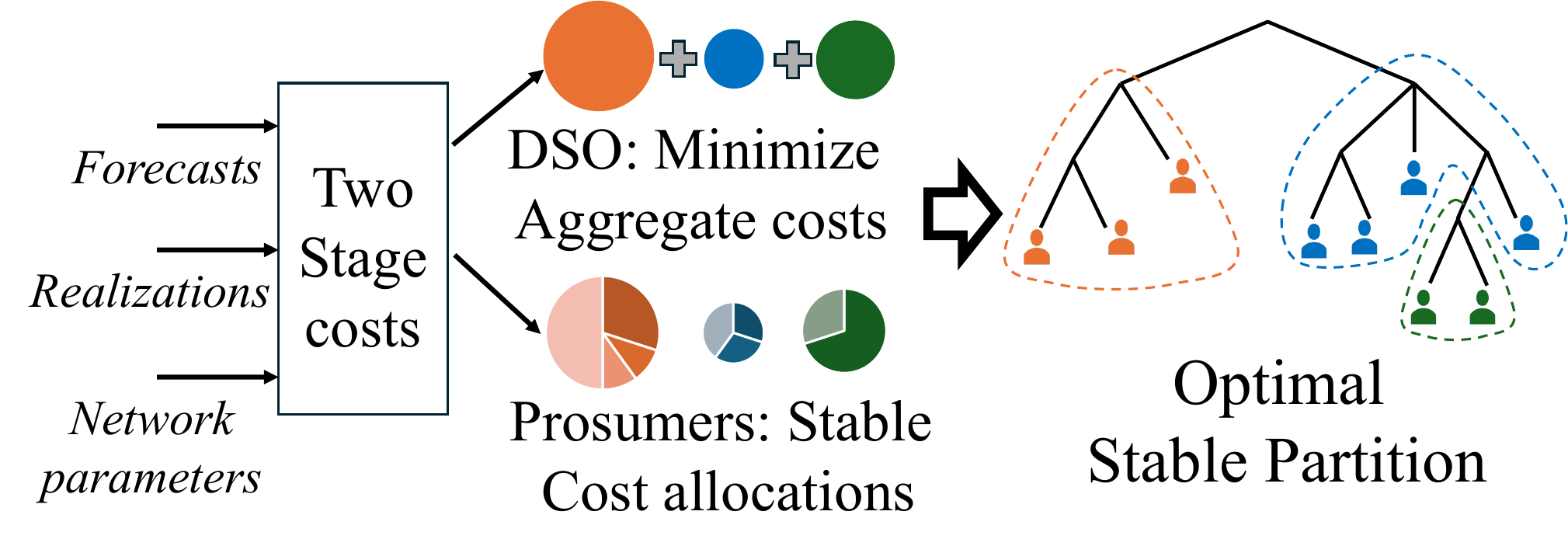}
    \caption{Optimal stable partition under forecasts and realizations}
    \label{fig:summaryfull}
\end{figure}

We denote the minimizer as $P^{\star}$ and refer to it as the optimal stable partition. This minimizes the operating costs (DSO's perspective), while constrained to partitions that are stable (prosumers' perspective). This is summarized in Fig \ref{fig:summaryfull}.


In general, the optimal stable partition depends on the grid parameters, prosumer configuration and prosumption uncertainty. Under the special case of coalition costs with no externalities (Eq \eqref{eq:coalitioncostsnoexternalities}) and perfect forecasts, one can find the solution for the optimal stable partition. 

\begin{theorem}
    (Optimal stable partition under perfect forecasts) Let $P_{GC}:=\{F_{GC}\}=\{(\mathcal{N},\mathcal{E})\}$. Then, under perfect forecasts $\boldsymbol{\hat U}=\boldsymbol{U}$, and Assumptions \ref{assm:boundarynode}-\ref{assm:strictselfconsume}, $P_{GC}$ is the optimal stable partition (Problem \eqref{eq:optimalstablepartition})
    \label{thm:stabilitystrictselfconsum}
\end{theorem}
We provide the proof in Appendix \ref{app:theorem1}. The theorem implies that under perfect forecasts and no externalities, not only is the largest LEM optimal, meaning it is preferred by the DSO (Proposition \ref{prop:optimalityperfectforecasts}), but it is also stable, that is, no set of prosumers would prefer forming a separate LEM.

We now consider the opposite extreme for grid partitioning, corresponding to each prosumer self-consuming individually. For the case of congested grids, we state the following result
\begin{proposition}
    (Optimal stable partition under grid congestion). Let $P_{singleton} = \{\{(m, m\rightarrow A_m)\}_{m\in\mathcal{M}_\mathcal{G}}\}$ be the partition that corresponds to singleton coalitions, that is, individual self-consumption. Then, under line impedances $z_n\rightarrow\infty \ \forall n\in\mathcal{N}$, power limits $\overline{S}_{n\rightarrow A_n} = 0 \ \forall n\in\mathcal{N}$, and Assumption \ref{assm:strictselfconsume}, $P_{singleton}$ is the optimal stable partition
    \label{prop:stabilitycongested}
\end{proposition}
The proof is provided in the Appendix \ref{app:thm2}. The two results establish conditions for stability and optimality for the two extreme cases of grid partitioning: Theorem \ref{thm:stabilitystrictselfconsum} shows that under perfect forecasts and no externalities, both the DSO and prosumers prefer forming the largest LEM, while Proposition \ref{prop:stabilitycongested} shows that for highly congested grids, the smallest LEMs (singleton coalitions) are preferred.

Under imperfect forecasts, moderate levels of grid congestion and/or coalition costs under externalities, the solution for the optimal stable partition (Problem \eqref{eq:optimalstablepartition}) depends on the grid configuration, cost parameters (flexibility, imbalance, voltage violations and line overloading penalties) and prosumption profiles and forecast error, as shown in Examples \ref{example:1} and \ref{example:2}. For this general case, we provide Algorithm \ref{alg:alg1} to determine the optimal stable partition.

\begin{algorithm}
\footnotesize
\caption{\small Optimal stable partitioning of the distribution grid}
\begin{algorithmic}[1]
\State $\mathcal{P}_{stable} \gets [\ ]$, \quad $\Phi_{stable} \gets [\ ]$
\For{$P \in \mathcal{P}(\mathcal{G})$}\label{line:exhaustivesearch}
    \State Compute $\Phi(P)$, $\boldsymbol{\lambda}, \boldsymbol{\alpha}^{\Delta E}$
    \State $ctr\gets 0$
    \For{$F_i\in P$}
        \If{Assumptions \ref{assm:boundarynode}-\ref{assm:strictselfconsume} are valid} \Comment{No Externalities}
            \State\label{line:noextercost} Compute $\phi(F_i;P)$ using \eqref{eq:coalitioncostsnoexternalities}
                \For{$F\in \mathcal{S}(F_i)$}
                \State\label{line:noexterdev} Compute deviation costs $\hat \phi(F)$ using Case \ref{case:noexter}
                \EndFor
        \EndIf
        \If{Assumptions \ref{assm:boundarynode}-\ref{assm:strictselfconsume} are not valid} \Comment{Externalities}
            \State\label{line:extercost} Compute $\phi(F_i;P)$ using \eqref{eq:coalitioncostsexternalities}
                \For{$F\in \mathcal{S}(F_i)$}
                \State\label{line:exterdev} Compute deviation costs $\hat \phi(F)$ using Case \ref{case:exter}
                \EndFor
        \EndIf
        \State\label{line:corelinearprogram} \[
            Y = \min_{\gamma}\ \sum_{j\in\mathcal{E}_{F_i}} \gamma_j 
            \quad
            \text{s.t. } \sum_{j\in\mathcal{E}_F}\gamma_j \leq \hat \phi(F)\ \forall F\in \mathcal{S}(F_i)
            \]
        \If{$Y = \phi(F_i;P)$} \Comment{Core is non-empty}
            \State $ctr\gets ctr+1$
        \EndIf
    \EndFor
    \If{$ctr==|P|$} \Comment{All coalitions are stable}
        \State Append $P$ to $\mathcal{P}_{stable}$ \Comment{Partition is stable}
        \State Append $\Phi(P)$ to $\Phi_{stable}$
    \EndIf
\EndFor
\State $P^\star \gets \arg\min_{P\in\mathcal{P}_{stable}}\ \Phi_{stable}$ \Comment{Minimize over stable partitions}
\end{algorithmic}
\label{alg:alg1}
\end{algorithm}

The algorithm takes the prosumption forecasts, realizations and grid configuration as input and outputs the optimal stable partition. This is done via an exhaustive search over the set of partitions (line \ref{line:exhaustivesearch}), to calculate aggregate partition costs. For each partition, coalition costs are then calculated, depending on the absence or presence of externalities (line \ref{line:noextercost} or line \ref{line:extercost}). Also conditioned on the externalities, deviation costs $\hat{\phi}(F)$ are calculated (line \ref{line:noexterdev} or line \ref{line:exterdev}) for each subcoalition of a coalition. After obtaining the deviation costs, the non-emptiness of the \emph{core} (\ref{def:coregeneral}) is checked by solving a linear program in line \ref{line:corelinearprogram}. If each of the coalitions is stable, the partition is added to the set of stable partitions. The optimal stable partition is calculated by finding the stable partition that minimizes aggregate costs.

Currently, Algorithm \ref{alg:alg1} is computationally prohibitive for the following reasons. Firstly, the algorithm performs an exhaustive search over the set of partitions (line \ref{line:exhaustivesearch}), which scales as the Bell number ($O((\frac{0.792|\mathcal{N}|}{\text{log}(|\mathcal{N}|+1)})^{|\mathcal{N}|})$ \cite{berend2010improved}). Secondly, checking for core non-emptiness involves a linear program (line \ref{line:corelinearprogram}), whose constraints scale as $O(2^{|\mathcal{M}_{F_i}|})$ based on the number of prosumers $|\mathcal{M}_{F_i}|$ in the coalition $F_i$. Putting these two together leads to a worst case scaling bound of $O((\frac{2\times0.792|\mathcal{N}|}{\text{log}(|\mathcal{N}|+1)})^{|\mathcal{N}|})$ for calculating the optimal stable partition. However, the focus of this work was on calculating exact solutions, so we stick to Algorithm \ref{alg:alg1} despite its poor scaling. We discuss extensions based on heuristic and approximation algorithms in Section \ref{sec:discussion}.

\section{Numerical Experiments}
\label{sec:numerics}
We now validate our theory through numerical experiments in two settings. First, we analyze stable partitioning under strict self-consumption requirement and varying levels of forecast noise in a modified IEEE 33 bus system. Following this, we calculate stable partitions in a model of an urban low-voltage grid in the city of Lausanne, Switzerland. 
\subsection{Partitioning under no externalities}
\label{sec:numericsieee}

\begin{figure}[h!]
    \centering
    \includegraphics[width=0.8\linewidth]{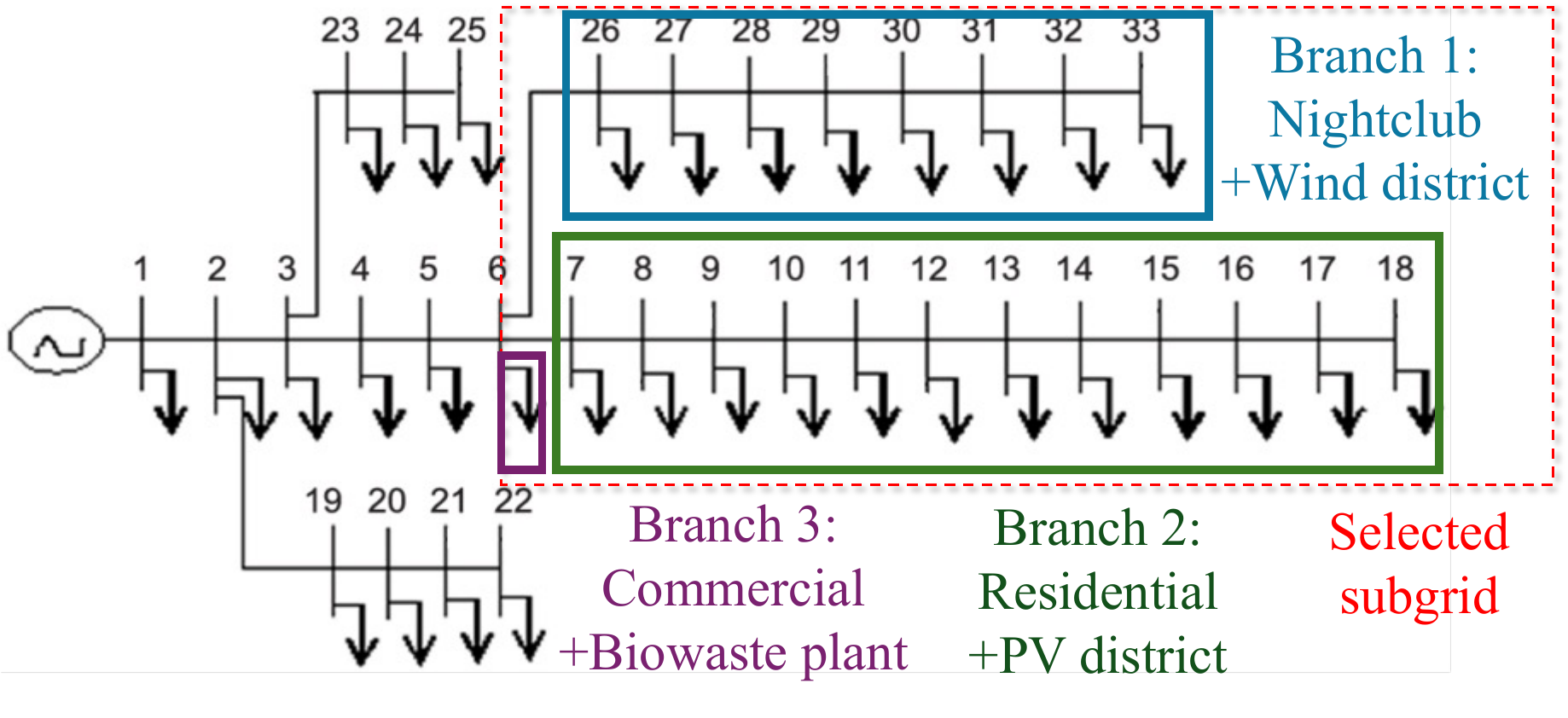}
    \caption{Modified IEEE 33 bus sytem}
    \label{fig:modifiedieee}
\end{figure}

We now study grid partitioning \eqref{eq:optimalstablepartition} under no externalities under Assumptions \ref{assm:boundarynode}-\ref{assm:strictselfconsume} using numerical experiments on a modified IEEE 33 bus system \cite{baran_optimal_1989}. For our experiments, we consider a reduced grid as shown in Fig \ref{fig:modifiedieee} where Node 6 is taken as the PCC of reduced grid which comprises of nodes 26-30, nodes 7-18 and the load at node 6. 

The modified distribution grid has three branches, and we let each branch comprise of prosumers of the same type: Prosumers in Branch 1 (nodes 26–33) have loads peaking at midnight and having stochastic (Rayleigh) wind generation; Branch 2 (nodes 7–18) prosumers are residential loads with rooftop PV; and Branch 3 (node 6) is a commercial load powered by a municipally funded waste incinerator generator.

\begin{figure}[h!]
    \centering

    \begin{minipage}[c]{0.48\linewidth}
        \centering
        \includegraphics[width=\linewidth]{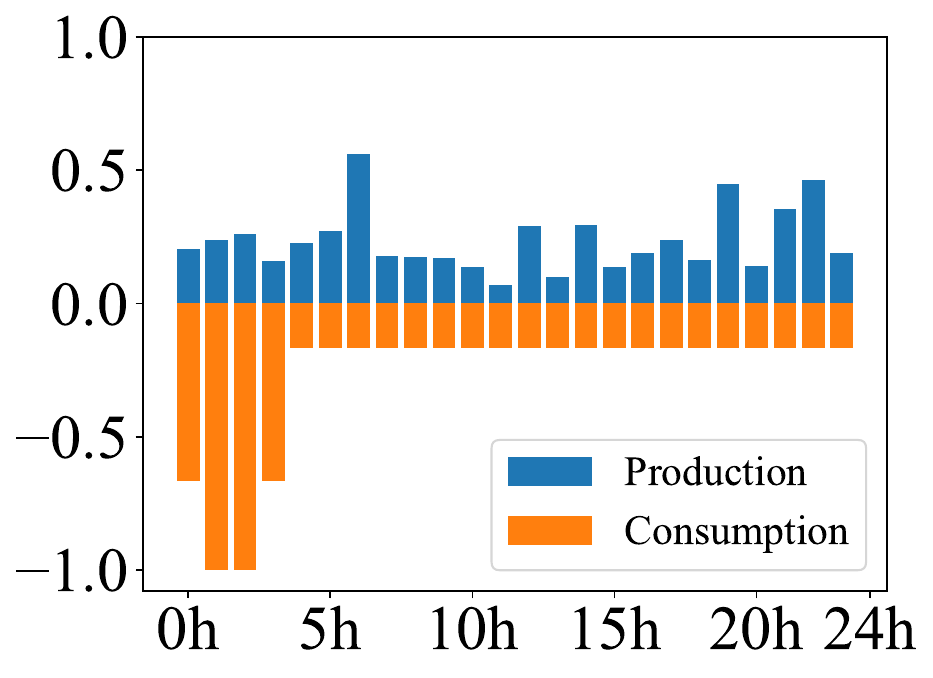}
        \caption*{(a) Branch 1 Prosumption profile 24h (p.u.)}
    \end{minipage}
    \hfill
    \begin{minipage}[c]{0.48\linewidth}
        \centering
        \includegraphics[width=\linewidth]{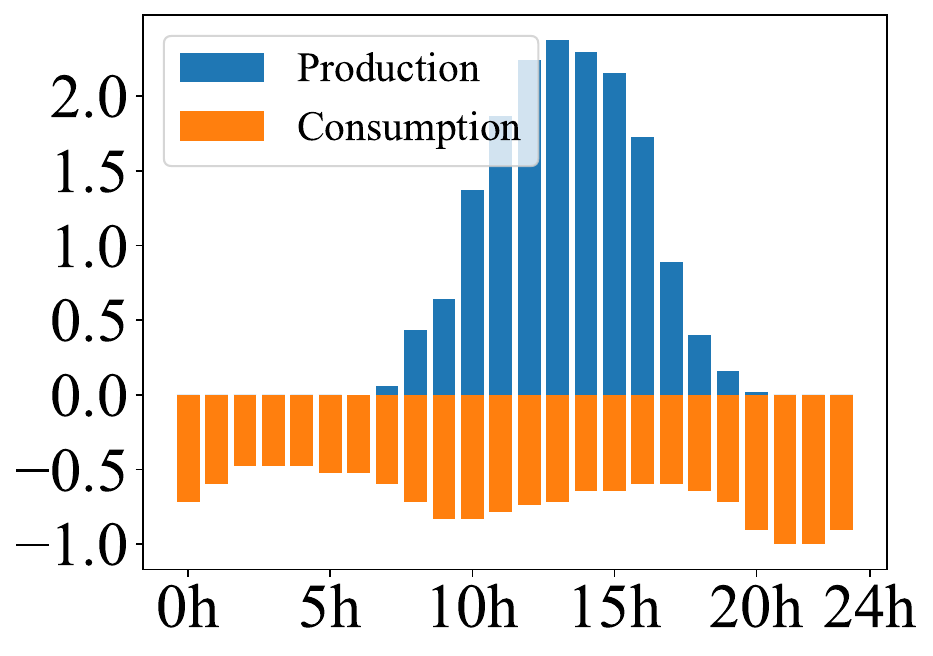}
        \caption*{(b) Branch 2 Prosumption profile 24h (p.u.)}
    \end{minipage}

    \vspace{1em} 

    \begin{minipage}[c]{0.48\linewidth}
        \centering
        \includegraphics[width=\linewidth]{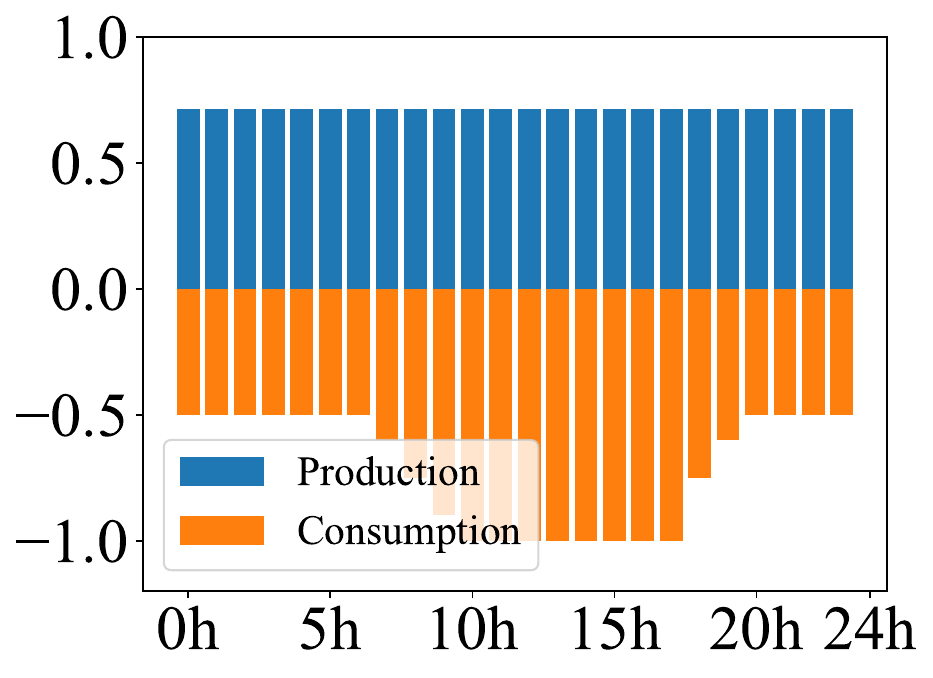}
        \caption*{(c) Branch 3 Prosumption profile 24h (p.u.)}
    \end{minipage}
    \hfill
    \begin{minipage}[c]{0.48\linewidth}
        \centering
        {\footnotesize
        \begin{tabular}{|c|c|}
            \hline
            Parameter & Cost \\
            \hline
            ESS 1 & $16$ CHF/MWh \\
            \hline
            ESS 2 & $24$ CHF/MWh \\
            \hline
            ESS 3 & $8$ CHF/MWh \\
            \hline
            Balancing & $300$ CHF/MWh \\
            \hline
        \end{tabular}
        \caption*{(d) Cost coefficients}}
    \end{minipage}

    \caption{Prosumption profiles and cost coefficients}
    \label{fig:profilesandcoeffs}
\end{figure}

Generation and load profiles for each of the three prosumer types are shown in Fig \ref{fig:profilesandcoeffs} (a), (b), (c), in peak units (p.u.) of load\footnote{To simulate grid overloading, peak consumption for branch 1 was scaled $1.6\times$, and branch 2 by $1.2\times$}. Generation capacities were chosen to ensure feasibility of individual self-consumption by each prosumer, with all generation connected behind the load without reactive power control. Each prosumer was equipped with an Energy Storage System (ESS), with operation costs given in Table \ref{fig:profilesandcoeffs} (d). 


To assess stable partitioning, Monte Carlo simulations were performed on 1000 variants of the previously described profiles, using random timestep scaling of up to 20\% and random shifts of up to 1 hour to reflect daily variability. Imperfect \emph{ex-ante} forecasts were generated by applying multiplicative noise of varying intensity to nodal prosumptions. \emph{Ex-post} voltage-violation penalties at each node were calculated using the costs for active power load curtailment, with curtailment costs assumed to be equal to imbalance costs given in Table \ref{fig:profilesandcoeffs} (d). 


\begin{table}[h!]
    \centering
    \begin{tabular}{|c|c|c|c|c|}
        \hline
        \multicolumn{1}{|c|}{} & \multicolumn{4}{c|}{Forecast Noise level} \\
        \hline
        Partitions & $0\%$ & $5\%$ & $10\%$ & $20\%$ \\
        \hline
        $\{\{1\},\{2\},\{3\}\}$ & $779$& $1097$ & $1413$& $\mathbf{2058}^*$\\
        \hline
        $\{\{1,2\},\{3\}\}$ & $504$ & $1164$ & $1925$ & $3156$ \\
        \hline
        $\{\{1,3\},\{2\}\}$ & $712$ & $1066$ & $\mathbf{1400}^*$ & $2065$ \\
        \hline
        $\{\{2,3\},\{1\}\}$ & $341$ & $1058$ & $1878$ & $3223$ \\
        \hline
        $\{\{1,2,3\}\}$ & $\mathbf{249}^*$ & $\mathbf{944}^*$ & $1723$ & $2975$ \\
        \hline
    \end{tabular}
    \caption{Total costs (CHF) for different forecast noise levels for each partition. \textbf{Boldface} indicates optimal partition, $^*$ indicates optimal stable partition}
    \label{tab:totalcostvspartitionnoise}
\end{table}

For simplicity, we only consider grid partitions that involve combinations of the three branches, and ignore partitions of prosumers within each branch. Note that under this setting, Assumption \ref{assm:boundarynode} holds for any subset of the three branches, and Assumption \ref{assm:strictselfconsume} is ensured by selection of energy storage and generation capacities as mentioned before. Hence this setting corresponds to partitioning under no externalities.

\begin{figure}[h!]
    \centering
    \includegraphics[width=0.8\linewidth]{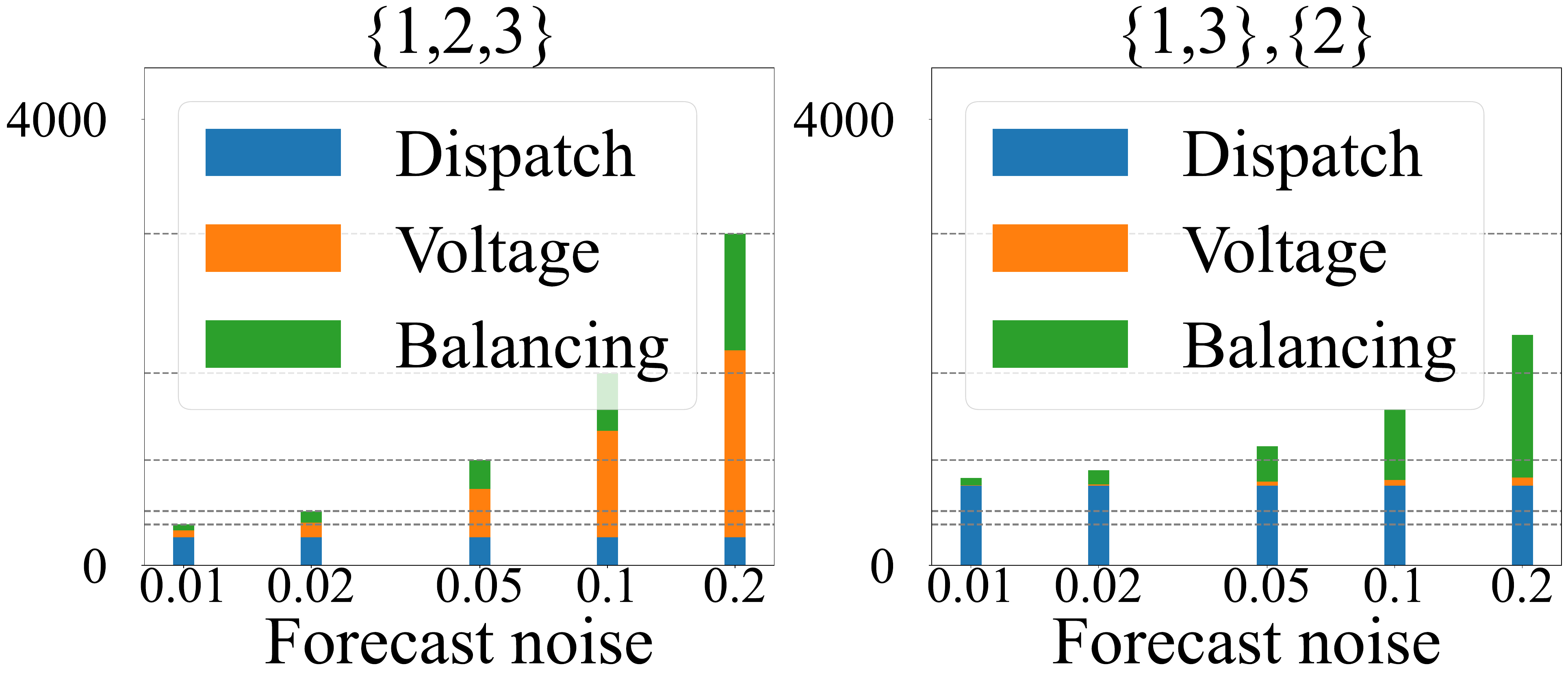}
    \caption{Cost breakup for two partitions for different noise levels: Voltage violation costs increase faster in larger markets}
    \label{fig:figcomparisonpartition}
\end{figure}

Aggregate costs and optimal stable partitions under varying levels of forecast noise are summarized in Table \ref{tab:totalcostvspartitionnoise}. We observe that as the forecast noise increases, prosumers organize in smaller LEMs to avoid expensive voltage violations. A representative comparison is shown in Fig \ref{fig:figcomparisonpartition}, where the costs are compared for the partitions $\{\{1,2,3\}\}$ and $\{\{1,2\},\{3\}\}$. For zero forecast noise, the largest LEM is the optimal stable partition, as predicted by Theorem \ref{thm:stabilitystrictselfconsum}.



Keeping the same setup, we also test the effect of grid congestion on the optimal stable partition. We now fix the forecast noise at $10\%$, and scale the network impedances by a range of factors to simulate grid congestion. The results are summarized in Table \ref{tab:totalcostvspartitioncong}. One can observe that for high impedance scale factors, the optimal stable partition corresponds to individual self-consumption, as predicted by Proposition \ref{prop:stabilitycongested}.

\begin{table}[h!]
    \centering
    \begin{tabular}{|c|c|c|c|}
        \hline
        \multicolumn{1}{|c|}{} & \multicolumn{3}{c|}{Impedance scale factor} \\
        \hline
        Partitions & $0.1\times$ & $1\times$ & $5\times$ \\
        \hline
        $\{\{1\},\{2\},\{3\}\}$ & $1482$ & $1413$& $\mathbf{1513}^*$\\
        \hline
        $\{\{1,2\},\{3\}\}$ & $1071$ & $1925$ & $25924$ \\
        \hline
        $\{\{1,3\},\{2\}\}$ & $1407$ & $\mathbf{1400}^*$ & $3009$ \\
        \hline
        $\{\{2,3\},\{1\}\}$ & $1035$ & $1878$ & $25892$ \\
        \hline
        $\{\{1,2,3\}\}$  & $\mathbf{806}^*$ & $1723$ & $27215$ \\
        \hline
    \end{tabular}
    \caption{Total costs (CHF) for different congestion levels for each partition. \textbf{Boldface} indicates optimal partition, $^*$ indicates optimal stable partition}
    \label{tab:totalcostvspartitioncong}
\end{table}



\subsection{Partitioning under externalities}
\label{sec:numericslausanne}
We now analyze stable partitioning in a real-world grid, where Assumptions \ref{assm:boundarynode}-\ref{assm:strictselfconsume} may not hold. Our experiments use a digital twin of a low-voltage network ($400 V$) section within the city of Lausanne, Switzerland, obtained in collaboration with the DSO, Services industriels de Lausanne. 

The network consists of a total of 43 prosumers with residential loads, some of which include photovoltaic generation. For the case study, we simulate partitioning outcomes under the assumption that all of these prosumers electrify their heating using heat pumps, replacing their current oil/gas based systems, without any grid reinforcements from the DSO's side. 

Each node's load and generation time series were obtained from anonymized smart meter data for residential buildings \cite{lepour_renewable_nodate}, which was then appropriately scaled to match the statistically estimated peak values for each prosumer provided by the DSO. Building characteristics were taken using the models for residential buildings from the work \cite{girardin_gis-based_2012}. The time series data for external temperature was similarly taken from \cite{girardin_gis-based_2012}. 

Prosumers were assumed to utilize their heat pumps towards flexibility, with a comfort range of $20^\circ C-24^\circ C$ and default internal temperature set to $22^\circ C$. The energy tariffs $\boldsymbol{\lambda}_0, \kappa_t$ were set according to the DSO's current values \cite{webmasterlausannech_electricite_2017}\footnote{In deriving $\kappa_t$ we assume that internal energy transport is fully discounted}. Imbalance penalty was set as $300\  CHF/MWh$ \cite{swissgrid_balance_nodate} with voltage penalty rates for each node calculated using the imbalance costs at the corresponding node, similar to the previous section.

Forecasts for solar irradiance and external temperature were assumed to be perfect, with the only source of uncertainty coming from consumption. The forecasts for daily consumption for each prosumer were generated using one-day-lagged time series as predictors for the next day.

\begin{figure}[h!]
    \includegraphics[width=0.8\linewidth]{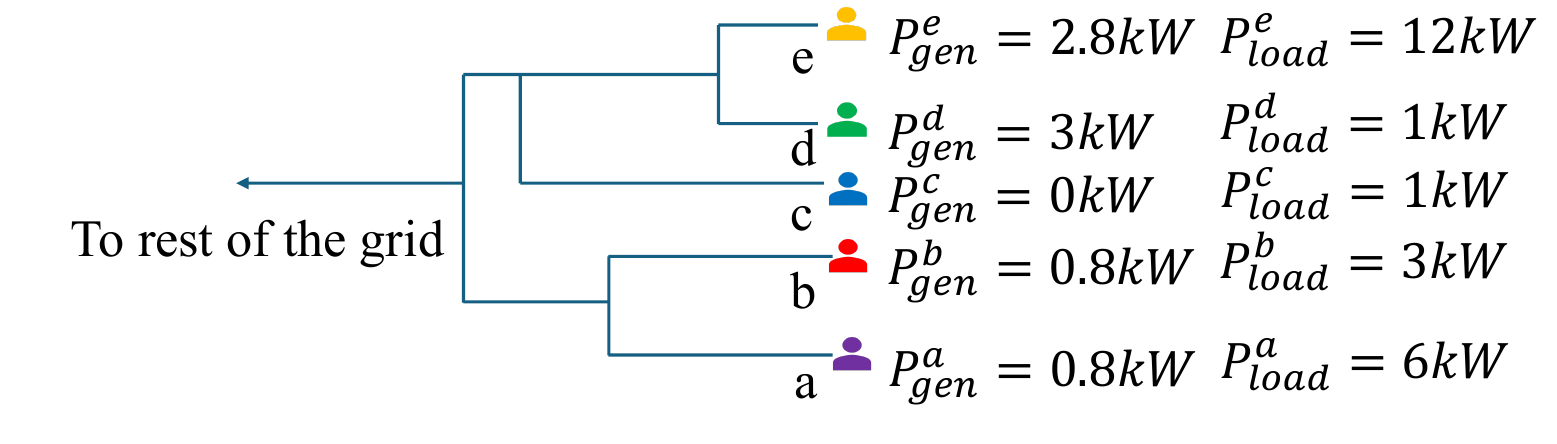} 
    \caption{Neighbourhood topology for LV grid section, with peak values for generation and load per prosumer}
    \label{fig:lausannepartition}
\end{figure}

Currently, in the city of Lausanne, each prosumer is given incentives to self-consume using tariffs \cite{webmasterlausannech_electricite_2017}, and there are no LEM coalitions. We take this as the baseline and investigate the partitioning outcomes when prosumers are allowed to form LEMs. While we used the prosumption and flexibility of all 43 prosumers for power flow analysis, for stable partitioning we only focus on a group of 5 prosumers located at the end of a feeder (Fig \ref{fig:lausannepartition}). The rest of the prosumers were assumed to not form coalitions and continue to operate as regular customers.

For these 5 prosumers, there were a total of 34 partitions that result in connected subgraphs. Each partition was checked for stability, using Algorithm \ref{alg:alg1}. The calculated costs for each of the possible partitions for this neighbourhood is given in the Appendix \ref{app:simulationslausanne}. In Table \ref{tab:lausannecosts} below, we summarize the results and compare the optimal stable partition against two other baseline partitions: \emph{1. A single collective LEM, 2. Individual self-consumption}.
\begin{table}[hbt!]
    \centering
    \begin{tabular}{|c|c|c|}
    \hline
         Partition & \makecell{Costs for\\ neighborhood} & \makecell{Costs for\\ entire network} \\
         \hline
         $\{\{a\},\{b\},\{c\},\{d\},\{e\}\}$ & 94 & 659 \\
         \hline
         $\mathbf{\{\{a, e\},\{b\},\{c\},\{d\}\}}$ & \textbf{57} & \textbf{582} \\
         \hline
         $\{\{a,b,c,d,e\}\}$ & 149 & 713 \\
         \hline
    \end{tabular}
    \caption{Costs (CHF) for different partitions}
    \label{tab:lausannecosts}
\end{table}

Table \ref{tab:lausannecosts} shows that the optimal stable partition differs from both baselines, indicating that prosumers prefer moderate level of localization over large coalitions or individual self-consumption. Furthermore, this partition minimizes costs for both the neighborhood and the distribution grid, suggesting that in this case, prosumer preferences align with the DSO's.

\subsection{Discussion}
\label{sec:discussion}
In the formulation, we had assumed that the LEM partitions remain fixed for a time horizon $\mathcal{T}$, which was an exogenous parameter. The next step would be to endogenize the selection of this time horizon. In principle, the optimal and stable partitions could change frequently with time in our current formulation, since prosumptions and grid conditions could vary across time. Practically, however, LEM formation comes with technical and organizational frictions, and these startup costs would need to be added to our current model for deviation costs $\hat{\phi}(F)$. Addition of these startup costs to coalitional deviation costs would relax the core constraints (Definition \ref{def:coregeneral}-\ref{def:stablepartition}), increase the likelihood the partition is stable, and discourage LEMs from changing over short time horizons. At the same time, for long horizons, these startup costs would be amortized, and partitions would again become more likely to change across time if grid conditions evolve. An important future extension for our work would be to find the relevant timescales for LEM formation under startup costs, using the theory of dynamic coalition formation \cite{ray_game-theoretic_2007}.

Regarding computation of optimal stable partition, the proposed algorithm performs an exhaustive search over the set of grid partitions, which is intractable for large grids. To extend the solution to practical use, one could use heuristics to reduce the size of the search space. Promising techniques include hierarchical spectral clustering for the network \cite{sanchez-garcia_hierarchical_2014}, Kron power flow reduction for the grid \cite{dorfler_kron_2013}, and graph neural network approaches \cite{shah_neurocut_2024}.

\section{Conclusion}
We formulated the optimal stable partitioning problem to study distribution grid partitioning under uncertain prosumption, from the perspective of the DSO and the prosumers. Under no uncertainty and strict self-consumption requirement, the largest LEM was shown to be the optimal stable partition (Theorem \ref{thm:stabilitystrictselfconsum}). In congested distribution grids, individual self-consumption was shown to be the optimal stable partition (Proposition \ref{prop:stabilitycongested}). For the general case of stochastic prosumption and grid congestion, we provided an algorithm to calculate this partition (Algorithm \ref{alg:alg1}).

Numerical results show that in constrained grids, finer LEM partitions are preferred by both the DSO and the prosumers as the uncertainty in prosumption increases. This was also validated by a small case study on a real world grid in Lausanne. On the other hand, as prosumption forecasts improve in the future, we predict that both the DSO and prosumers may prefer larger LEMs, as indicated by Theorem \ref{thm:stabilitystrictselfconsum} and Section \ref{sec:numericsieee}.

Future work could investigate efficient algorithms for partitioning, incorporating power losses, fairness and privacy considerations in LEM formation, extending to unbalanced networks, as well as joint analysis of investment and operation decisions in partitioning.  
\label{sec:concl}
\section*{Acknowledgements}

We thank Max Chevron (Services industriels de Lausanne) for the data support related to Lausanne grid. We thank Maitraya Desai for valuable comments related to the manuscript.

\appendices

\section{Proofs}
\renewcommand{\thesubsection}{\thesection.\arabic{subsection}}
\renewcommand{\thesubsectiondis}{\thesection.\arabic{subsection}}
\renewcommand{\theequation}{\thesection.\arabic{equation}}
\setcounter{equation}{0}

\subsection{Proof for Proposition \ref{prop:costscoalitionnoexternalities}}\label{app:prop1}
For $\kappa_t\rightarrow\infty$, Eq \eqref{eq:coalitioncostsexternalities} only has terms for flexibility costs, costs for constraint violations in its internal network, and imbalance costs. We show that these reduce to Eq \eqref{eq:coalitioncostsnoexternalities}.

Firstly, we show that power flow constraints can be decoupled. Since each LEM has exactly one boundary node (Assumption \ref{assm:boundarynode}), each LEM is a radial graph with all of its nodes as descendants of its boundary node. Hence, all coalitions can be treated as leaf nodes to get the transformed radial graph $\mathcal{G}^{\prime}$ from the original $\mathcal{G}$. Since each LEM deploys flexibility and reserves for zero power exports $\boldsymbol{S}_{PCC(F)}(\boldsymbol{u},\boldsymbol{U};F) = \boldsymbol{S}_{PCC(F)}(\boldsymbol{u},\boldsymbol{\hat U};F) = 0$ (Assumption \ref{assm:strictselfconsume}), there are no power flows in edges of $\mathcal{G}^{\prime}$, and thus $\boldsymbol{\delta v}_{PCC(F_i)} = 0\ \forall F_i\in P$. Hence the feasible set $\mathcal{B}_{\mathcal{G}}(\boldsymbol{\hat U})$ can be decomposed for each LEM as ($\mathcal{B}_F(\boldsymbol{\hat U})$, Eq \eqref{eq:boundaryzeroexchangecoalitionproposition}), and power flow constraints are decoupled for all $F_i\in P$. 


Since power flow constraints are decoupled, and \emph{ex-ante} objective (Eq \eqref{eq:firststagecosts}) is separable over LEMs, $\boldsymbol{u}_{F_i}^I$ (Eq \eqref{eq:exanteobjectivecoalitionproposition}) is equal to $\{\boldsymbol{u}_n^I\}_{n\in F}$ Eq (\eqref{eq:firststagesolution}). Hence, flexibility costs (Eq \eqref{eq:coalitioncostsexanteproposition}) correspond to flexibility costs in (Eq \eqref{eq:intcostscoalition}). 

\emph{Ex-post} costs for constraint violations in internal networks of LEMs (Eq \eqref{eq:lemconstraintviolations}) can be obtained using decoupled power flow and voltage equations $\boldsymbol{S}(\boldsymbol{u}_{F_i},\boldsymbol{U};F_i), \ \boldsymbol{\delta v}(\boldsymbol{u}_{F_i},\boldsymbol{U};F_i)$ in (Eq \eqref{eq:coalitioncostsinternalproposition}). As a result of independent energy balancing and zero power exports by LEMs (Assumption \ref{assm:strictselfconsume}), there are no costs for constraint violations in external nodes and edges \eqref{eq:commonconstraintviolations}, and energy balancing costs are given as Eq \eqref{eq:balancecostscoalitionsproposition}.

\subsection{Proof for Proposition \ref{prop:optimalityperfectforecasts}}\label{app:lemma1}
Under perfect forecasts, the \emph{ex-post} costs vanish, and the total costs are equal to the \emph{ex-ante} objective Eq \eqref{eq:firststagecosts}. We write these costs for a general partition $P\in\mathcal{P}(\mathcal{G})$, which are given as the minimizer for the ex-ante dispatch problem

\begin{equation}
\begin{aligned}
    \Phi(P) = \min_{\boldsymbol{u}}&\sum_{l=1}^L \sum_{m\in\mathcal{M}_{F_l}}c_m (\boldsymbol{u}_m)-\left\langle\boldsymbol{\lambda}_0,\boldsymbol{U}_m+\boldsymbol{u}_m\right\rangle \\
        +&\sum_{l=1}^L\sum_{t\in\mathcal{T}}\sum_{n_b\in\mathcal{N}_{F_i}^b}\kappa_t \left\lVert S^t_{n_b}(\boldsymbol{u},\boldsymbol{U};F_i,\mathcal{G}) \right\rVert_2\\
        &\mathrm{s.t.}\  \eqref{eq:voltagedroppartition}-\eqref{eq:feasibleflexpartition}
\end{aligned}
\label{eq:partitioncosttotalappendix}
\end{equation}
We now compare the above partition cost $\Phi(P)$ against the cost for the grid partition having a single LEM (grand coalition) $P = P_{GC}$. Notice that the set of feasible flexibility dispatches, specified by the constraints Eq \eqref{eq:voltagedroppartition}-\eqref{eq:feasibleflexpartition} is the same for all partitions. Additionally, note that the first two terms of the objective, which represent the flexibility costs and export revenue, are the same across all partitions. Only the penalty term from the tax rate on power exchanges differs across partitions. Hence, we only need to compare the penalty term for the grand coalition against any other partition.

The grand coalition only has one boundary node, which is also the PCC of the distribution grid. Thus, the penalty term can be written as 
$\sum_{t\in\mathcal{T}}\kappa_t \left\lVert S_{PCC(\mathcal{G})}^t(\boldsymbol{u},\boldsymbol{U})\right\rVert_2$. Notice that the power flows through the PCC of the distribution grid are equal to the total power exports of all prosumers. For given prosumptions $\boldsymbol{U}$ and flexibility dispatch $\boldsymbol{u}$, the total power exports $S_{PCC(\mathcal{G})}^t(\boldsymbol{u},\boldsymbol{U})$ can be written as the sum of power exports of each LEM $\sum_{i=1}^L\sum_{n_b\in\mathcal{N}_{F_i}^b}{S}^t_{n_b}(\boldsymbol{u},\boldsymbol{U};F_i)$ in any given partition $P=\{F_i\}_{i=1}^L$.

Using Jensen's inequality, $\forall P=\{F_i\}_{i=1}^L\in\mathcal{P}(\mathcal{G}):$
\begin{align}
\begin{aligned}
    &\sum_{t\in\mathcal{T}}\sum_{i=1}^L\sum_{n_b\in\mathcal{N}_{F_i}^b}\left\lVert{S}^t_{n_b}(\boldsymbol{u},\boldsymbol{U};F_i)\right\rVert_2\\
    \ & \geq \sum_{t\in\mathcal{T}}\left\lVert\sum_{i=1}^L\sum_{n_b\in\mathcal{N}_{F_i}^b}{S}^t_{n_b}(\boldsymbol{u},\boldsymbol{U};F_i)\right\rVert_2\\
    \ &= \sum_{t\in\mathcal{T}}\kappa_t \left\lVert S_{PCC(\mathcal{G})}^t(\boldsymbol{u},\boldsymbol{U})\right\rVert_2
\end{aligned}
\end{align}
Thus the penalty term is minimized for the grand coalition\footnote{Note that in our work, we have modelled self-consumption incentives via adding a penalty on magnitude of power flows (L2-norm), to keep it consistent with existing tariff structures in Lausanne \cite{webmasterlausannech_electricite_2017}. Similar results should hold for other convex penalty terms (such as a lower injection price compared to consumption price), but we avoid this in our work for brevity.}. Since the other terms in the optimization objective are the same, and the feasible set is the same across all partitions, the minimizer of the problem \eqref{eq:partitioncosttotalappendix} cannot have a higher value for the grand coalition compared to any other partition, and hence $\Phi(P_{GC})\leq\phi(P) \ \forall P\in\mathcal{P}(\mathcal{G})$ under perfect forecasts.




\subsection{Proof for Theorem \ref{thm:stabilitystrictselfconsum}}\label{app:theorem1}
Under perfect forecasts, the total costs for the partition $P_{GC}=\{F_{GC}\}=\{\mathcal{N}_{\mathcal{G}},\mathcal{E}_{\mathcal{G}}\}$ are given by 
\begin{align}
\begin{aligned}
    &\Phi(P_{GC})
    =\min_{\boldsymbol{u}}\ \sum_{n\in\mathcal{M}_{\mathcal{G}}} c_n(\boldsymbol{u}_n)\\
    &\qquad\qquad \mathrm{s.t.\ } \boldsymbol{S}_{PCC(\mathcal{G})}(\boldsymbol{u},\boldsymbol{U}) \overset{(A.1a)}{=} 0, \eqref{eq:voltagedroppartition}-\eqref{eq:feasibleflexpartition}
\label{eq:appcgcpartitioncost}
\end{aligned}
\end{align}
To show Theorem \ref{thm:stabilitystrictselfconsum}, it suffices to show that $P_{GC}=\{F_{GC}\}$ is the optimal partition (shown in Proposition \ref{prop:optimalityperfectforecasts}) and that it is a stable partition. For the latter, we find one cost allocation that lies in its \emph{Core} (Def \ref{def:coregeneral})\footnote{Partition stability for the grand coalition $\{F_{GC}\}$ involves showing non-emptiness of the core. Prior work \cite{azim_dynamic_2024} claims this in a related setting using an incorrect argument: they show superadditivity and monotonicity (inequalities (43), (44) in their paper), but incorrectly subtract these to claim supermodularity. We instead show non-emptiness by constructing an explicit allocation.}. 

We will use the Lagrangian corresponding to the optimization problem \eqref{eq:appcgcpartitioncost}. Denote the dual variables for the constraints \eqref{eq:voltagedroppartition} as $\boldsymbol{\beta}_n$, \eqref{eq:nodalpowerbalancepartition} as $\boldsymbol{\lambda}_n$, \eqref{eq:forwardpowerlimitpartition} as $\boldsymbol{\mu}_n$, \eqref{eq:voltagelimitpartition} as $\boldsymbol{\eta}_n$, and \eqref{eq:feasibleflexpartition} as $\theta_n$. Finally, denote the dual variables for \eqref{eq:refvoltagegridpartition} as $\boldsymbol{\beta}_0$ and for $A.1b$ as $\boldsymbol{\lambda}_0$. 

Let $\Psi=\{ \boldsymbol{S},\boldsymbol{\delta v},\boldsymbol{u}\}$ compactly denote the primal variables, and denote the dual variables as $\Lambda=\{\boldsymbol{\beta},\boldsymbol{\lambda}, \boldsymbol{\eta},\boldsymbol{\mu},\theta\}$. The total costs \eqref{eq:appcgcpartitioncost} can be written using the Lagrangian $\mathcal{L}(\Psi,\Lambda)$ as
\begin{align}
\begin{aligned}
    \Phi(P_{GC}) = \phi(F_{GC};P_{GC}) = \mathcal{L}(\Psi^{opt},\Lambda^{opt})
\end{aligned}
\end{align}
where $\Psi^{opt}, \Lambda^{opt}$ are the primal and dual optimizers respectively. Now, define $\mathcal{L}_n(\Psi_n, \Lambda)$ as the following
\begin{enumerate}
        \item For $n=PCC(\mathcal{G})$, $\Psi_0=\{\boldsymbol{\delta v}_{PCC(F)},\boldsymbol{S}_{PCC(F)}\}$. 
        \begin{align*}
            \mathrm{Let\ }\mathcal{L}_0(\Psi_0,\Lambda) =& \left\langle\sum_{m\in D_{PCC(\mathcal{G})}}\boldsymbol{\beta}_m-\boldsymbol{\beta}_0, \boldsymbol{\delta v}_{PCC(\mathcal{G})}\right\rangle\\
            &+\left\langle \boldsymbol{\lambda}_0, \boldsymbol{S}_{PCC(\mathcal{G})}\right\rangle
        \end{align*}
        \item For $n\in\mathcal{N}\setminus PCC(\mathcal{G})$, define $\Psi_n = \{\boldsymbol{S}_{e_n}, \boldsymbol{\delta v}_n\}$. Denote
        \begin{align*}
        \mathcal{L}_n(\Psi_n,\Lambda) 
        &=\  \Omega_n(\Psi_n, \Lambda)+\left\langle\sum_{m\in D_n}\boldsymbol{\beta}_m-\boldsymbol{\beta}_n, \boldsymbol{\delta v}_n\right\rangle\\
        &+2\mathrm{Re}(\left\langle z_n\boldsymbol{\beta}_n, \boldsymbol{S}_{e_n}\right\rangle)+\left\langle \boldsymbol{\lambda}_n-\boldsymbol{\lambda}_{A_n}, \boldsymbol{S}_{e_n}\right\rangle\\
        &+\left\langle\boldsymbol{\mu}_n,|\boldsymbol{S}_{e_n}|^2-\overline{S}_{e_n}^2\right\rangle+\left\langle \boldsymbol{\eta}_n, |\boldsymbol{\delta v}_n|-\overline{\delta v_n}\right\rangle
        \end{align*}
        with $\Psi_n = \{\boldsymbol{u}_n,\boldsymbol{S}_{e_n}, \boldsymbol{\delta v}_n\}$ if $n\in\mathcal{M}_{\mathcal{G}}$ otherwise $\Psi_n=\{\boldsymbol{S}_{e_n}, \boldsymbol{\delta v}_n\}$. $\Omega_n(\Psi_n, \Lambda) = c_n(\boldsymbol{u}_n)+\theta_n g_n(\boldsymbol{u}_n)-\left\langle\boldsymbol{\lambda}_n,\boldsymbol{U}_n+\boldsymbol{u}_n\right\rangle$ if $n\in\mathcal{M}_{\mathcal{G}}$ and 0 otherwise.
\end{enumerate}

One can show that $\sum_{n\in\mathcal{N}_{\mathcal{G}}}\mathcal{L}_n(\Psi_n,\Lambda) = \mathcal{L}(\Psi,\Lambda)$, that is, the Lagrangian is of a separable form with respect to the primal variables $\Psi_n$. As a result of this, we obtain for $\ \forall n\in\mathcal{N}_{\mathcal{G}}$

\begin{align}
    \argmin_{\Psi_n} \mathcal{L}_n(\Psi_n,\Lambda^{opt}) = & \argmin_{\Psi_n} \mathcal{L}(\Psi,\Lambda^{opt})
    \label{eq:separablelagrangemin}
\end{align}
Now, choose the cost allocation over edges as $\boldsymbol{y}\in\mathbb{R}^{|\mathcal{E}_{\mathcal{G}}|}$, with $\boldsymbol{y}_{(n\rightarrow A_n)} = \mathcal{L}_n(\Psi_n^{opt},\Lambda_n^{opt})$. Firstly, applying primal feasibility for the equations \eqref{eq:refvoltagegridpartition} and $A.1a$, one obtains $\mathcal{L}_0(\Psi_0,\Lambda)=0$. As a result, we satisfy budget balance:
\begin{align*}
\sum_{e\in\mathcal{E}_{F_{GC}}}\boldsymbol{y}_e = \sum_{n\in\mathcal{N}\setminus PCC(\mathcal{G})}\mathcal{L}_n(\Psi_n^{opt},\Lambda^{opt}) = \phi(F_{GC};P_{GC})
\end{align*}
The allocation $\sum_{i\in\mathcal{E}_F}\boldsymbol{y}_i$ for any defecting LEM $F$ satisfying Assumption \ref{assm:boundarynode} can be written as

\begin{align*}
\begin{aligned}
    &\sum_{e\in \mathcal{E}_F} \boldsymbol{y}_e = \sum_{n\in\mathcal{N}_F\setminus PCC(F)}\mathcal{L}_n(\Psi_n^{opt},\Lambda^{opt})
\end{aligned}
\end{align*}
Using the above definitions for $\mathcal{L}_n(\Psi_n,\Lambda)$, Karush-Kuhn-Tucker conditions for \eqref{eq:voltagedroppartition}-\eqref{eq:feasibleflexpartition} for the convex problem Eq \eqref{eq:appcgcpartitioncost}, separability of Lagrangian \eqref{eq:separablelagrangemin}, and that $F$ is a radial subgraph with one PCC (Assumption \ref{assm:boundarynode}), this simplifies to


\begin{align}
    \begin{aligned}
    &\min_{\{\Psi_n\}_{n\in \mathcal{N}_F^{int}}} \ \sum_{n\in \mathcal{M}_F}c_n(\boldsymbol{u}_n)-\left\langle\boldsymbol{\lambda}_{PCC(F)}^{opt},\sum_{m\in D_{PCC(F)}}\boldsymbol{S}_{e_m} \right\rangle\\
    &\qquad-\left\langle\sum_{m\in D_{PCC(F)}}\boldsymbol{\beta}_{m}^{opt},\boldsymbol{\delta v}_{m}-2\mathrm{Re}(z_m\boldsymbol{S}_{e_m})\right\rangle\\
    &\qquad\mathrm{s.t.}\ \eqref{eq:voltagedroppartition}-\eqref{eq:feasibleflexpartition}
    \end{aligned}
    \label{eq:allocationcoalition}
\end{align}
If one adds constraints $\sum_{m\in D_{PCC(F)}}\boldsymbol{S}_{e_m} = 0$, $\boldsymbol{\delta v}_{PCC(F)}=\sum_{m\in D_{PCC(F)}}\boldsymbol{\delta v}_{m}-2\mathrm{Re}(z_m\boldsymbol{S}_{e_m}) = 0$, \eqref{eq:allocationcoalition} becomes
\begin{align}
\begin{aligned}
    &\min_{\boldsymbol{u}_F\in\mathcal{B}_{F}(\boldsymbol{U})} \sum_{n\in \mathcal{M}_F}c_n((\boldsymbol{u}_F)_n)\\
    &\mathrm{s.t.}\ \boldsymbol{S}_{PCC(F)}(\boldsymbol{u},\boldsymbol{U};F) = 0
    \end{aligned}
    \label{eq:allocationcoalitionconstraints}
\end{align}

\eqref{eq:allocationcoalitionconstraints} is equal to $\hat \phi(F) = \phi_0(F)$ under perfect forecasts, that is, the costs for the deviating coalition $F$ satsifying Assumptions \ref{assm:boundarynode}-\ref{assm:strictselfconsume}. Since addition of constraints cannot yield a lower value for \eqref{eq:allocationcoalition}, we conclude that $\sum_{e\in\mathcal{E}_F}\boldsymbol{y}_e\leq \hat \phi(F)$
\subsection{Proof for Proposition \ref{prop:stabilitycongested}}
\label{app:thm2}
Firstly, via Assumption \ref{assm:strictselfconsume}, note that the feasibility of individual self-consumption is ensured, even for the relaxed case of finite $\kappa_t$. For the case of $z_n\rightarrow\infty, \overline{S}_{n\rightarrow A_n}=0 \ \forall n\in\mathcal{N}$, from the power flow equations Eq \eqref{eq:voltagedroppartition}-\eqref{eq:feasibleflexpartition}, one can observe that the only feasible solution to the ex-ante dispatch problem is $\boldsymbol{u_m}^I=-\hat{\boldsymbol{U}}_m \ \forall m\in\mathcal{M}_G$, with each prosumer $m\in\mathcal{M}_\mathcal{G}$ independently balancing the energy imbalance of  $\hat{U}_m^t-U_m^t \ \forall t\in\mathcal{T}$ in the ex-post stage.

As a result, note that for any partition $P=\{F_l\}_{l=1}^L\in\mathcal{P}(\mathcal{G})$, the boundary power flows $S^t_{n_b}(\boldsymbol{u},\boldsymbol{\hat U};F_l,\mathcal{G})=0\ \forall n_b\in\mathcal{N}_{F_l}^b, \ \forall F_l\in P$. Similarly for each prosumer $m\in\mathcal{M}_\mathcal{G}$ the power exports $\hat{\boldsymbol{U}}_m+\boldsymbol{u}_m$ are zero. Since there is only one feasible solution for the energy dispatch problem across all partitions, the two stage cost for all partitions $P\in\mathcal{P}(\mathcal{G})$ are equal, and given by the following
\begin{align*}
    \Phi(P;\hat{\boldsymbol{U}},\boldsymbol{U}) = &\sum_{m\in\mathcal{M}_\mathcal{G}}\left(c_m(-\hat{\boldsymbol{U}}_m)\right.\\
    &+\left.\sum_{t\in\mathcal{T}}\alpha_{(m, m\rightarrow A_m)}^{\Delta E} \|\hat{U}_m^t-U_m^t\|_2\right)
\end{align*}
Thus, the partition corresponding to singleton coalitions $\{\{(m, m\rightarrow A_m)\}_{m\in\mathcal{M}_\mathcal{G}}\}$ attains the optimum value, and is the optimal partition for the DSO (Problem \eqref{eq:optimalpartitiondso}).

The two stage cost for any coalition in the distribution grid $F\in\mathcal{S}(\mathcal{G})$ is given by
\begin{align*}
    \phi_0(F) = &\sum_{m\in\mathcal{M}_F}\left(c_m(-\hat{\boldsymbol{U}}_m)\right.\\
    &+\left.\sum_{t\in\mathcal{T}}\alpha_{(m, m\rightarrow A_m)}^{\Delta E} \left\lVert\hat{U}_m^t-U_m^t\right\rVert_2\right)
\end{align*}
One observes that the coalition cost is additive on the set of prosumers. Thus each prosumer will get the same allocation equal to $c_m(-\hat{\boldsymbol{U}}_m)+\sum_{t\in\mathcal{T}}\|\hat{U}_m^t-U_m^t\|_2$, regardless of its coalition membership, and every coalition is trivially stable. Thus the partition $\{\{(m, m\rightarrow A_m)\}_{m\in\mathcal{M}_\mathcal{G}}\}$ corresponding to singleton LEMs is also stable. Since it was shown to be the optimal partition, it is also the optimal stable partition.


\section{Simulation details}
\label{app:simulations}
\subsection{IEEE 33 bus system}
\label{app:simulationsiee}
The simulation was carried out on MacBook Pro 2023 (Apple M3 CPU and RAM of 18GB) using a computer program written in Python. The experiment consisted of a Monte Carlo experiment calculating the two stage coalition and partition costs over 1000 randomisations of prosumption profiles of a representative day. This was analyzed for varying forecast noise levels $\{0.0,0.01,0.02,0.05,0.1,0.2\}$. This experiment required a run time of 17 hours and 16 minutes. To simulate and grid congestion levels, network impedances were scaled by a factor of $\{0.1\times, 1\times\, 5\times\}$, keeping the forecast noise fixed at $10\%$. This experiment required a run time of 8 hours and 20 minutes. \\
The code is available on Github: \url{https://github.com/saurabh-vaishampayan/Optimal-Stable-Partitioning-of-Distribution-Grids.git}

\subsection{Lausanne LV grid}
\label{app:simulationslausanne}

The simulation was carried out on MacBook Pro 2023 (Apple M3 CPU and RAM of 18GB) using a computer program written in Python, with the total time of 2 hours and 45 minutes required for simulation. To keep the discussion brief in the main text, in Table \ref{tab:lausannecosts} we had only provided a summary of the calculation results. We now provide the calculated costs for all of the 34 partitions of the neighbourhood in Table \ref{tab:lausannecostsfull}.

\begin{table}[hbt!]
    \centering
    \begin{tabular}{|c|c|c|c|}
    \hline
         Partition& \makecell{Coalition costs\\(CHF)} & \makecell{Partition\\ costs\\ (CHF)} & \makecell{Network\\costs\\(CHF)} \\ 
         \hline
         \{\{a\}, \{b\}, \{c\}, \{d\}, \{e\}\} & \{21,  8, 43,  2, 21\} & 95 & 660 \\
        \hline
        \{\{b, a\}, \{c\}, \{d\}, \{e\}\} & \{32, 43,  3, 21\} & 99 & 662 \\
        \hline
        \{\{b\}, \{c, a\}, \{d\}, \{e\}\} & \{10, 90, 0, 36\} & 135 & 703 \\
        \hline
        \{\{b\}, \{c\}, \{d, a\}, \{e\}\} & \{10, 11, 25, 36\} & 81 & 637 \\
        \hline
        \textbf{\{\{b\}, \{c\}, \{d\}, \{e, a\}\}}$^*$ & \textbf{\{10,  6,  2, 39\}} & \textbf{58} & \textbf{582} \\
        \hline
        \{\{a\}, \{c, b\}, \{d\}, \{e\}\} & \{23, 74, -1, 36\} & 132 & 681 \\
        \hline
        \{\{c, b, a\}, \{d\}, \{e\}\} & \{94,  0, 35\} & 128 & 668 \\
        \hline
        \{\{a\}, \{c\}, \{d, b\}, \{e\}\} & \{23,  7, 11, 36\} & 78 & 620 \\
        \hline
        \{\{c\}, \{d, b, a\}, \{e\}\} & \{10, 33, 35\} & 78 & 618 \\
        \hline
        \{\{a\}, \{c\}, \{d\}, \{e, b\}\} & \{24, 10,  2, 27\} & 62 & 582 \\
        \hline
        \{\{c\}, \{d\}, \{e, b, a\}\} & \{34,  3, 50\} & 87 & 610 \\
        \hline
        \{\{a\}, \{b\}, \{d, c\}, \{e\}\} & \{18,  6, 70, 36\} & 130 & 700 \\
        \hline
        \{\{b, a\}, \{d, c\}, \{e\}\} & \{27, 70, 36\} & 133 & 701 \\
        \hline
        \{\{b\}, \{d, c, a\}, \{e\}\} & \{ 10, 102,  36\} & 148 & 716 \\
        \hline
        \{\{a\}, \{d, c, b\}, \{e\}\} & \{23, 84, 36\} & 143 & 693 \\
        \hline
        \{\{d, c, b, a\}, \{e\}\} & \{103,  35\} & 138 & 676 \\
        \hline
        \{\{a\}, \{b\}, \{d\}, \{e, c\}\} & \{ 21,   8,   3, 100\} & 132 & 698 \\
        \hline
        \{\{b, a\}, \{d\}, \{e, c\}\} & \{ 33,   3, 101\} & 136 & 700 \\
        \hline
        \{\{b\}, \{d\}, \{e, c, a\}\} & \{ 10,   3, 112\} & 125 & 650 \\
        \hline
        \{\{a\}, \{d\}, \{e, c, b\}\} & \{ 24,   2, 109\} & 135 & 658 \\
        \hline
        \{\{d\}, \{e, c, b, a\}\} & \{  3, 135\} & 138 & 700 \\
        \hline
        \{\{a\}, \{b\}, \{c\}, \{e, d\}\} & \{21,  8, 45, 32\} & 105 & 674 \\
        \hline
        \{\{b, a\}, \{c\}, \{e, d\}\} & \{33, 45, 32\} & 109 & 676 \\
        \hline
        \{\{b\}, \{c, a\}, \{e, d\}\} & \{10, 90, 37\} & 136 & 705 \\
        \hline
        \{\{b\}, \{c\}, \{e, d, a\}\} & \{10,  6, 50\} & 67 & 590 \\
        \hline
        \{\{a\}, \{c, b\}, \{e, d\}\} & \{23, 74, 38\} & 134 & 683 \\
        \hline
        \{\{c, b, a\}, \{e, d\}\} & \{94, 36\} & 129 & 668 \\
        \hline
        \{\{a\}, \{c\}, \{e, d, b\}\} & \{24, 10, 37\} & 71 & 591 \\
        \hline
        \{\{c\}, \{e, d, b, a\}\} & \{34, 61\} & 95 & 619 \\
        \hline
        \{\{a\}, \{b\}, \{e, d, c\}\} & \{ 22,   8, 113\} & 142 & 711 \\
        \hline
        \{\{b, a\}, \{e, d, c\}\} & \{ 33, 114\} & 147 & 713 \\
        \hline
        \{\{b\}, \{e, d, c, a\}\} & \{ 10, 125\} & 135 & 659 \\
        \hline
        \{\{a\}, \{e, d, c, b\}\} & \{ 24, 120\} & 144 & 667 \\
        \hline
        \{\{e, d, c, b, a\}\} & \{149\} & 149 & 713 \\
        \hline
    \end{tabular}
    \caption{Two stage daily operation costs for Lausanne LV grid for all partitions and coalitions, averaged over one month. \textbf{Boldface} indicates optimal partition, $^*$ indicates optimal stable partition}
    \label{tab:lausannecostsfull}
\end{table}
As mentioned in Section \ref{sec:numericslausanne}, in this case the optimal partition (\{\{a,e\},\{b\},\{c\},\{d\}\}, highlighted in \textbf{bold}) also turned out to be a stable partition and thus also the optimal stable partition.

Additionally, from Table \ref{tab:lausannecostsfull}, one can also observe the presence of coalitional externalities, where a coalition's costs not only depend on its internal configuration, but also on the coalitions of prosumers external to it. In the partition $\{\{b,a\},\{d,c\},\{e\}\}$, the coalition $\{b,a\}$ obtains a cost of $27$ CHF, while in the partition $\{\{b,a\},\{e,d,c\}\}$, it gets a cost of $33$ CHF.

\printbibliography

\end{document}